\documentclass[a4paper,twocolumn,superscriptaddress,11pt,accepted=2019-05-28]{quantumarticle}
\pdfoutput=1
\usepackage[utf8]{inputenc}
\usepackage[english]{babel}
\usepackage[T1]{fontenc}
\usepackage{amsmath,amssymb,braket}
\usepackage{hyperref}
\usepackage[numbers,sort&compress]{natbib}

\usepackage{tikz}
\usepackage{lipsum}

\begin{document}

\title{A Quantum N-Queens Solver}

\author{Valentin Torggler}
\affiliation{Institute for Theoretical Physics, University of Innsbruck, A-6020
Innsbruck, Austria}

\author{Philipp Aumann}
\affiliation{Institute for Theoretical Physics, University of Innsbruck, A-6020
Innsbruck, Austria}

\author{Helmut Ritsch}
\affiliation{Institute for Theoretical Physics, University of Innsbruck, A-6020
Innsbruck, Austria}

\author{Wolfgang Lechner}

\email{wolfgang.lechner@uibk.ac.at}
\affiliation{Institute for Theoretical Physics, University of Innsbruck, A-6020
Innsbruck, Austria}
\affiliation{Institute for Quantum Optics and Quantum Information of the Austrian
Academy of Sciences, A-6020 Innsbruck, Austria}

\date{\today}
\begin{abstract}
The $N$-queens problem is to find the position of $N$ queens on an $N$ by $N$ chess board such that no queens attack each other. The excluded diagonals $N$-queens problem is a variation where queens cannot be placed on some predefined fields along diagonals. This variation is proven NP-complete and the parameter regime to generate hard instances that are intractable with current classical algorithms is known. We propose a special purpose quantum simulator that implements the excluded diagonals $N$-queens completion problem using atoms in an optical lattice and cavity-mediated long-range interactions. Our implementation has no overhead from the embedding allowing to directly probe for a possible quantum advantage in near term devices for optimization problems.
\end{abstract}



\maketitle

\section{Introduction}

Quantum technology with its current rapid advances in number, quality and controllability of quantum bits (qubits) is approaching a new era with computational quantum advantage for numerical tasks in reach \cite{harrow2017quantum,cirac2012goals,blatt2012quantum,RevBloch,aspuru2012photonic,de2018accurate,houck2012chip,RevNori,saffman2016quantum,preskill2018quantum,boixo2018characterizing}. While building a universal gate-based quantum computer with error-correction is a long-term goal, the requirements on control and fidelity to perform algorithms with such a universal device that outperform their classical counterparts are still elusive. Building special purpose quantum computers with near-term technology and proving computational advantage compared to classical algorithms is thus a goal of the physics community world wide \cite{preskill2012quantum}. Quantum simulation with the aim to solve Hamiltonian systems may serve as a building block of such a special purpose quantum computer \cite{bernien2017probing,kim2010quantum,schauss2012observation}. In particular, adiabatic quantum computing \cite{Nishimori1998,farhi2000quantum,albash2016adiabatic} has been proposed to solve computationally hard problems by finding the ground state of Ising spin glasses \cite{lucas2014ising}. Despite considerable theoretical \cite{albash2016adiabatic} and experimental \cite{boixo2014evidence} efforts, quantum speedup in adiabatic quantum computing has not been demonstrated in an experiment yet \cite{hauke2019perspectives}. Thus, demonstrating quantum advantage by solving optimization problems using  quantum simulation tools is a crucial step towards the development of general programmable quantum optimizers \cite{lechner2015,choi2008}.

\begin{figure}[tb]
  \centering
  \includegraphics[width=\columnwidth]{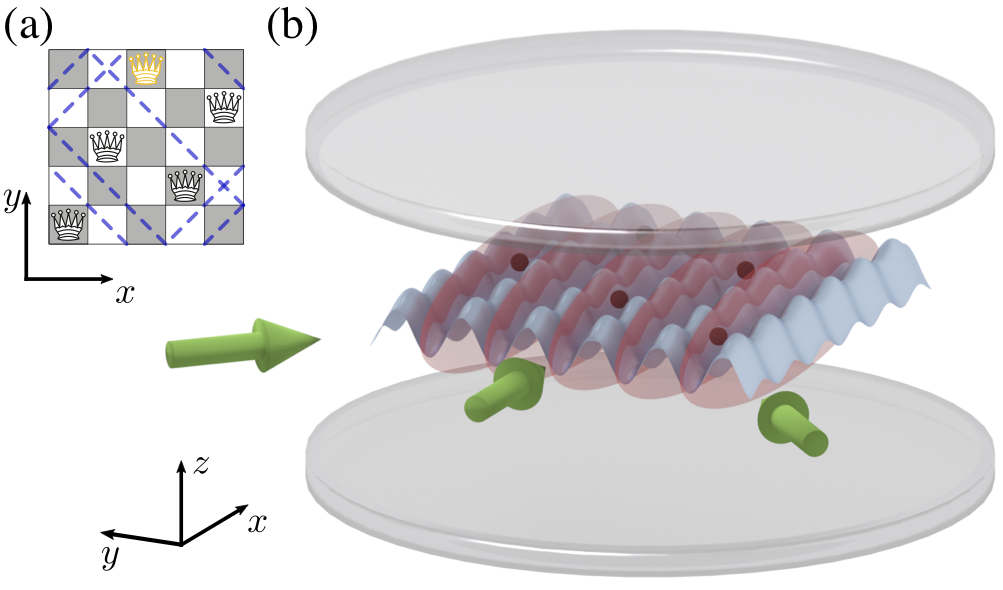}
  \caption{\textit{Sketch of the setup.} - (a) The $N$-queens problem is to place $N$ non-attacking queens on an $N$ by $N$ board. A variation thereof is the N-queens completion problem where some of the queens are already placed (yellow). In addition, some excluded diagonals are introduced (dashed blue lines) on which no queen can be placed. (b) Each queen is represented by an atom which is trapped in an anisotropic optical potential (blue) allowing for tunneling in $x$-direction only. Collective scattering of pump laser light (green arrows) into an optical resonator induces atom-atom interactions, preventing atoms from aligning along the $y$-axis and along the diagonals. After initial preparation in superposition states delocalized in $x$-direction (red tubes), increasing the interactions transfers the system into its solid phase, which is the solution of the queens problem (black balls).}
  \label{fig:fig1}
\end{figure}

Here we present a scheme that aims at solving the $N$-queens problem, and variations of it, using atoms with cavity-mediated long-range interactions \cite{ritsch2013cold,maschler2005cold,landig2016quantum,mekhov2012quantum,caballero2016quantum}.
We note that the $N$-queens problem is not just of mathematical interest but also has some applications in computer science \cite{bell2009survey}. In this work, variations of the problem are used as a testbed \cite{gent2017complexity} to study a possible quantum advantage in solving classical combinatorial problems in near term quantum experiments.

Our proposed setup consists of $N$ ultracold atoms in an optical lattice that represent the queens on the chess board \cite{hen2017realizable}. The non-attacking conditions are enforced by a combination of restricted hopping \cite{hen2016quantum} and interactions between the atoms stemming from collective scattering of pump laser light into a multi-mode cavity \cite{domokos2002collective,black2003observation,vaidya2018tunable,gupta2007cavity,gopalakrishnan2011frustration,gopalakrishnan2012exploring,kramer2014self,torggler2017quantum} (see Fig.\ \ref{fig:fig1}). For the excluded diagonals variation of the $N$-queens problem, additional repulsive optical potentials are introduced. The solution of the problem (or the ground state of the many-body quantum system) is attained via a superfluid-to-solid transition. From the measurement of photons that leave the cavity \cite{mekhov2009quantum} it can be determined if a state is a solution to the $N$-queens problem. The position of the atoms can in addition be read out with single site resolved measurement. The final solution is a classical configuration and thus easy to verify. We show that a full quantum description of the dynamics is required to find this solution.

Following Ref.\ \cite{harrow2017quantum}, we identify a combination of several unique features of the proposed model that makes it a viable candidate to test quantum advantage in near term devices. (a) The completion and excluded diagonals problem is proven to be NP-complete and hard instances for the excluded diagonals variant are known from computer science literature \cite{gent2017complexity}, (b) the problem maps naturally to the available toolbox of atoms in cavities and thus can be implemented without intermediate embedding and no qubit overhead, (c) the verification is computationally simple and (d) the number of qubits required to solve problems which are hard for classical computers ($N > 21$ for the solvers used in Ref.\ \cite{gent2017complexity}) is available in the lab.

Methods such as minor embedding \cite{choi2008,choi2011}, LHZ \cite{lechner2015,rocchetto2016stabilizers,glaetzle2017coherent} or nested embedding \cite{vinci2018scalable} always cause a qubit overhead. Here the intermediate step of embedding the optimization problem in an Ising model is removed by implementing the infinite-range interactions with cavity-mediated forces tailored to the problem's geometry in combination with constrained tunneling \cite{hen2016quantum}. Hence there is no qubit overhead and the mode resources scale linearly with $N$. The required number of qubits is reduced from several hundreds to below 50, which is available in current experiments. By implementing our scheme with less than 50 atoms the problem is already hard to tackle with current classical algorithms \cite{gent2017complexity}.

Light-mediated coupled tunneling gives rise to non-local quantum fluctuations across the whole lattice in the intermediate stage of the transition \cite{hormozi2017nonstoquastic,albash2019role}. Their non-uniform signs stemming from the relative phases of the cavity fields from site to site indicate that the system's Hamiltonian is non-stoquastic and can thus not be efficiently simulated with path integral Monte Carlo methods on a classical computer \cite{isakov2016understanding}. The question of a quantum speed-up is thus open unless a local transformation to a stoquastic Hamiltonian is found \cite{klassen2019twolocal,marvian2019computational}.

In our model implementation the non-local qubit interactions are mediated via the field modes of an optical resonator, which will attain non-classical atom-field superposition states during the parameter sweep. This appears to be an essential asset of the system as we find that the ground-state is reached only with a very low probability, when the full quantum dynamics of the fields is replaced by a classical mean-field approximation.

As a final feature let us point out here that the verification of a solution is computationally trivial as the final state is classical and no quantum tomography is needed. In principle the convergence to a solution can be simply deduced from the cavity outputs at the end of the sweep. With this, the proposed setup may serve as a platform to demonstrate combinatorial quantum advantage in near-term experiments.

This work is organized as follows: In Sec.\ \ref{sec:model} we introduce a quantum model based on coupled quantum harmonic oscillators simulating the $N$ queens problem. A proposed physical implementation using ultracold atoms in optical lattices and light-mediated atom-atom interaction is described in Sec.\ \ref{sec:implementation}. In Sec.\ \ref{sec:justification} we present a numerical comparison between model and implementation including photon loss. Finally we discuss in Sec.\ \ref{sec:readout} how light leaking out of the cavity can be used for read-out and we conclude in Sec.\ \ref{sec:conclusions}.

\section{Quantum simulation of the $N$-queens problem}
\label{sec:model}

Following the idea of adiabatic quantum computation \cite{Nishimori1998,farhi2000quantum,albash2016adiabatic}, we construct a classical problem Hamiltonian $H_\mathrm{pr}$ such that its ground state corresponds to the solutions of the $N$-queens problem. In order to find this ground state, the system is evolved with the time-dependent Hamiltonian
\begin{equation}
\label{eq:sweep}
H(t) = H_\mathrm{kin} + \frac{t}{\tau} H_\mathrm{pr}
\end{equation}
from $t=0$ to $t=\tau$. Initially at $t=0$, the system is prepared in the ground state of $H(0)=H_\mathrm{kin}$. During the time evolution, the second term is slowly switched on. If this parameter sweep is slow enough, the system stays in the instantaneous ground state and finally assumes the ground state of $H(\tau)$ at $t=\tau$. If the lowest energy gap of $H_\mathrm{pr}$ is much larger than the one of $H_\mathrm{kin}$, this state is close to the ground state of $H_\mathrm{pr}$ and thus the solution of the optimization problem.

In the following we construct the problem Hamiltonian $H_\mathrm{pr}$ and the driver Hamiltonian $H_\mathrm{kin}$. The system is modeled as a 2D Bose-Hubbard model with annihilation (creation) operators $b_{ij}$ ($b_{ij}^\dagger$) on the sites $(i,j)$. A position of a queen is represented by the position of an atom in an optical lattice with the total number of atoms being fixed to $N$. The non-attacking condition between queens, which amounts to interactions between two sites $(i,j)$ and $(k,l)$, is implemented with four constraints: There can not be two queens on the same line along (i) the $x$-direction $j=l$, (ii) the $y$-direction $i = k$, the diagonals (iii) $i+j = k+l$ and (iv) $i-j = k- l$.

Condition (i) is implemented by using an initial state with one atom in each horizontal line at $y_j$ and restricting the atomic movement to the $x$-direction [see Fig. \ref{fig:fig1}(b)]. Thereby we use the a priori knowledge that a solution has one queen in a row, which reduces the accessible configuration space size from ${N^2}\choose{N}$ to $N^N$ configurations. In this vein the restricted tunneling Hamiltonian \cite{hen2016quantum} is given by
\begin{equation}
\label{eq:kineticHamiltonian}
H_\mathrm{kin} = -  J \sum_{i,j=1}^N \hat B_{ij},
\end{equation}
where $J$ is the tunneling amplitude and $\hat B_{ij} = b_{i,j}^{\dagger}  b_{i+1,j} +  b_{i+1,j}^{\dagger} b_{i,j}$ with $\hat B_{Nj} = 0$ are the tunneling operators.

Constraints (ii), (iii) and (iv) are enforced by infinite range interactions between the atoms with 
\begin{equation}\label{eq:HQ}
H_\mathrm{Q} = U_\mathrm{Q} \sum_{ijkl =1}^{N} A_{ijkl} \ \hat n_{ij} \hat n_{kl},
\end{equation}
where $\hat n_{i,j}=b_{i,j}^{\dagger} b_{i,j}$ and $U_\mathrm{Q} > 0$. The interaction matrix is
\begin{equation}\label{eq:A}
A_{ijkl} =
	\begin{cases}
		3 & \text{if } (i,j) = (k,l) \\
		1 & \text{if } i=k \lor i+j = k+l \lor i-j = k-l  \\
		0 & \text{otherwise},
	\end{cases}
\end{equation}
where in the first case all three constraints are broken.

In order to implement variations of the $N$-queens problem, we need to exclude diagonals (for the excluded diagonals problem) and pin certain queens (for the completion problem). These additional conditions are implemented by local energy offsets of the desired lattice sites
\begin{equation}\label{eq:Hpot}
H_\mathrm{pot} = U_\mathrm{D} \sum_{i,j=1}^{N} D_{ij} \ \hat n_{ij} - U_\mathrm{T} \sum_{i,j=1}^{N} T_{ij} \ \hat n_{ij}.
\end{equation}
For $U_\mathrm{D} > 0$ the first term renders occupations of sites on chosen diagonals energetically unfavorable. Each diagonal (in $+$ and $-$ direction) has an index summarized in the sets $\mathcal{D}_+$ and $\mathcal{D}_-$, respectively, and the coefficients are
\begin{equation}\label{eq:D}
D_{ij} =
\begin{cases}
2 & \text{if } i+j-1 \in \mathcal{D}_+ \land  i-j+N \in \mathcal{D}_- \\
1 & \text{if } i+j-1 \in \mathcal{D}_+ \lor  i-j+N \in \mathcal{D}_-  \\
0 & \text{otherwise}.
\end{cases}
\end{equation}
For $U_\mathrm{T} > 0$, the second term favors occupations of certain sites. The sites where queens should be pinned to are pooled in the set $\mathcal{T}$ and therefore the coefficients are given by
\begin{equation}\label{eq:T}
T_{ij} =
\begin{cases}
1 & \text{if } (i,j) \in \mathcal{T} \\
0 & \text{otherwise}.
\end{cases}
\end{equation}
The problem Hamiltonian of the $N$-queens problem with excluded diagonals is then
\begin{equation}
\label{eq:Hpr}
H_\mathrm{pr} = H_\mathrm{Q} + H_\mathrm{pot}.
\end{equation}
Note that due to the initial condition atoms never meet and sites are occupied by zero or one atom only. Hence the system can be effectively described by spin operators \cite{hen2017realizable, torggler2017quantum}, also without large contact interactions.

\begin{figure}[tb]
	\centering
	\includegraphics[width=\columnwidth]{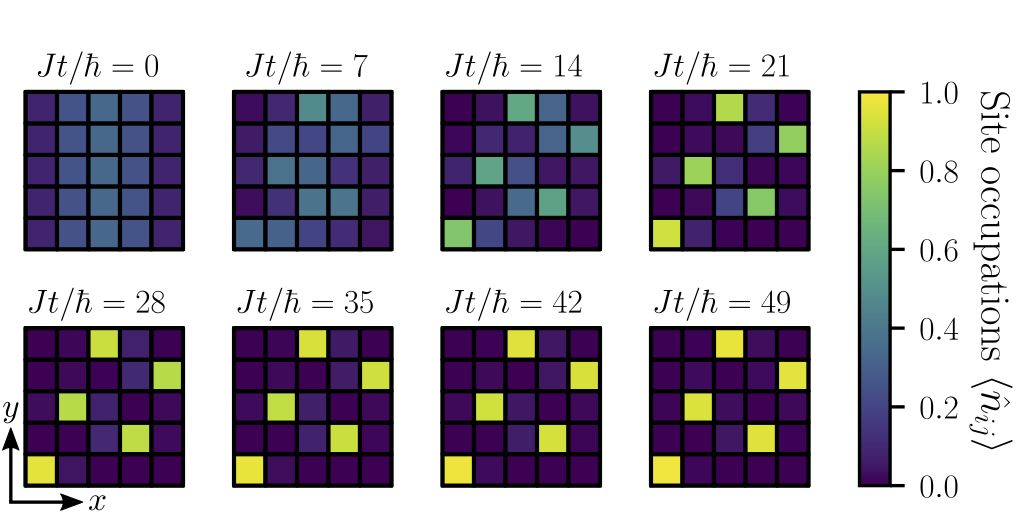}
	\caption{\textit{Time evolution of site occupations.} - Each subplot shows a snapshot of the site occupations $\langle \hat n_{ij} \rangle$ for the parameter sweep in Eq.\ \eqref{eq:sweep} and a sweep time $J\tau / \hbar = 49$. This is the instance shown in Fig.\ \ref{fig:fig1}a, where the excluded diagonals are indexed by $\mathcal{D}_+ = \{2,3,6,9\}$ and $\mathcal{D}_- = \{1,2,8,9\}$ and one queen is pinned at site $(3,5)$, i.e.\ $\mathcal{T}=\{(3,5)\}$ (see text). The final values of the sweep are $U_\mathrm{Q} = J$, $U_\mathrm{D} = 5J$, $U_\mathrm{T} = 2J$ while $J$ is kept constant. Since the sweep time is large enough, the state of the system adiabatically converges to the unique solution of the problem, which can be easily verified.}
	\label{fig:timeevolution}
\end{figure}

Let us illustrate the parameter sweep in Eq.\ \eqref{eq:sweep} for a specific example instance with $N=5$ queens (see Fig.\ \ref{fig:fig1}). The excluded diagonals chosen here restrict the ground state manifold to two solutions, and by biasing site $(3,5)$ one of these solutions is singled out. The time evolution of the site occupations $\langle \hat n_{ij} \rangle$ from numerically solving the time-dependent Schr\"odinger equation is shown in Fig.\ \ref{fig:timeevolution}. Initially, the atoms are spread out in $x$-direction since the ground state of $H(0)=H_\mathrm{kin}$ is a superposition of excitations along each tube. After evolving for a sufficiently large time $J \tau / \hbar = 49$, the system is in the ground state of $H_\mathrm{pr}$ and thus assumed the solution of the optimization problem.

\begin{figure}[htb]
  \centering
  \includegraphics[width=\columnwidth]{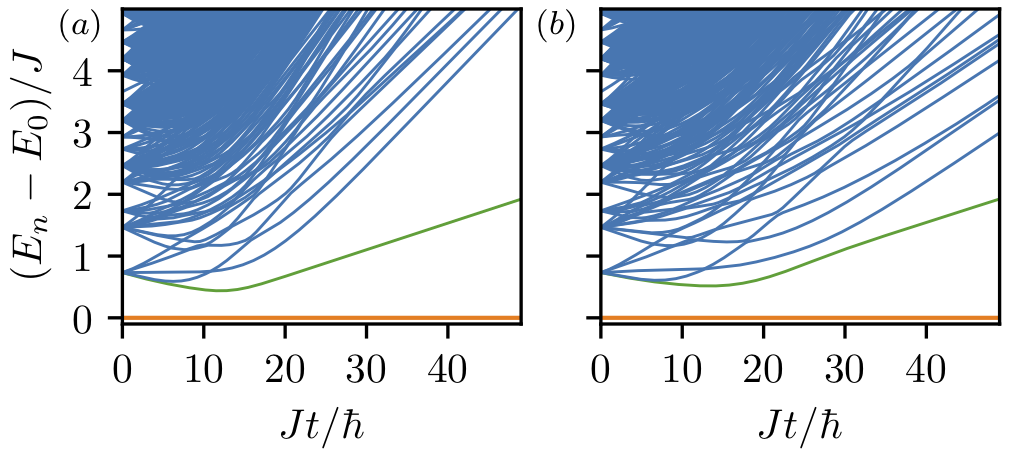}
  \caption{\textit{Energy spectrum.} - Eigenvalue spectrum of $H(t)$ for (a) the model Hamiltonian [Eq.\ \eqref{eq:sweep}] and (b) the Hamiltonian created by the field-atom interactions [Eq.\ \eqref{eq:sweepModes}] for the instance described in Fig.\ \ref{fig:timeevolution}. The orange and green lines show ground state and first excited state, respectively. The blue lines depict higher energy eigenstates. The low energy sector, and especially the minimal gap, are qualitatively similar.}
  \label{fig:spectrum}
\end{figure}

The energy spectrum of the given instance is shown in Fig.\ \ref{fig:spectrum}(a). The minimal gap between ground state (orange) and first excited state (green) determines the minimum sweep time $\tau$ to remain in the ground state according to the Landau-Zener formula. At the end of the sweep, the ground state closely resembles the solution to the excluded diagonals problem shown in Fig.\ \ref{fig:fig1}.

The Hilbert space for the atomic state corresponding to the configuration space mentioned above grows exponentially as $N^N$ and thus, as usual for quantum systems, the computational costs get large for rather small systems. Simulations with significantly larger systems are hence not easily tractable.

\section{Implementation}
\label{sec:implementation}

Here we propose a specific and at least conceptionally simple and straightforward physical implementation of Eq.\ \eqref{eq:sweep}, where the $N$ queens are directly represented by $N$ ultracold atoms in an $N$x$N$ two-dimensional optical lattice. Assuming tight binding conditions the atoms are confined to the lowest band of the lattice. They can coherently tunnel between sites \cite{bloch2005ultracold} and interact via collective light scattering within an optical resonator. To implement the required queens interactions via light scattering we make use of a set of optical field modes in a multi-mode standing-wave resonator [see Fig.\ \ref{fig:fig1}(b)]. It was shown before that this configuration in principle allows to implement arbitrary site to site interactions \cite{torggler2017quantum} if a sufficient number of modes is used.

To reduce the necessary Hilbert space without loss of generality, it is sufficient to enable tunneling only along one dimension ($x$), e.g. the rows of the lattice and increase the lattice depth in the column ($y$) direction. This confines the atoms (queens) to move only along tubes in the $x$-direction forming a parallel array of $N$ 1D optical lattices as routinely used to study 1D physics with cold atoms \cite{meinert2014observation}.

Luckily, for the special case of the infinite range queens interactions, one can find a strikingly simple and intuitive example configuration requiring only few field modes for each of the three remaining queens interaction directions. For this we consider the optical lattice to be placed in the center symmetry plane of the optical resonator, where one has a common anti-node of all symmetric eigenmodes. The trapped atoms are then illuminated by running plane wave laser beams from three different directions within the lattice plane with frequencies matched to different longitudinal cavity modes.

Depending on the size of the problem we need to add more laser frequencies in each direction to avoid a periodic recurrence of the interaction within the lattice. Note that all frequencies can be easily derived and simultaneously stabilized from a single frequency comb with a spacing matched to the cavity length. Since all light frequencies are well separated compared to the cavity line width, each is scattered into a distinct cavity mode and as the cavity modes are not directly coupled no relative phase stability is needed \cite{torggler2017quantum}. The resulting position-dependent collective scattering into the cavity then introduces the desired infinite-range interactions between the atoms. Choosing certain frequencies and varying their relative pump strengths allows for tailoring these interactions to simulate the three non-attacking conditions in the queens problem discussed in the previous section. Additional light sheets and local optical tweezers can be used to make certain diagonals energetically unfavorable (for the excluded diagonal problem) or pin certain queens. Alternatively pinning can be achieved by selective cavity mode injection. 

In the first part (Sec.\ \ref{ssec:tight-binding}) we will now introduce a Bose-Hubbard-type Hamiltonian for the trapped atoms interacting via collective scattering for general illumination fields in more detail. For this we first consider the full system of coupled atoms and cavity modes and later adiabatically eliminate the cavity fields to obtain effective light-mediated atom-atom interactions.

In the second part (Sec.\ \ref{subsec:nqueensinteraction}) we discuss how to implement the non-attacking conditions of the $N$-queens problem with such light-mediated atom-atom interactions. Specifically, we consider the limit of a deep optical lattice leading to simple analytical expressions. These formulas allow us to find a specific pump configuration of wave numbers and pump strengths leading to the ideal queens interaction Hamiltonian in Eq.\ \eqref{eq:HQ}. As a deep lattice depth slows down atomic tunneling it requires long annealing times. Luckily, it turns out that the pump configuration derived for a deep lattice still gives the same ground state for moderate lattice depth with faster tunneling. We show this by numerical simulations in Sec.\ \ref{sec:justification}. While the details of the interaction are altered it still sufficiently well approximates the $N$-queens interaction.

\subsection{Tight-binding model for atoms interacting via light}
\label{ssec:tight-binding}

Driving far from any atomic resonance the internal degrees of freedom of the atoms can be eliminated. In this so-called dispersive limit the resulting effective Hamiltonian couples the atomic motion to the light fields \cite{domokos2003mechanical}.

\textit{Single-particle Hamiltonian.} For a single particle of mass $m_A$, the motion of the atoms in the $x$-$y$-plane is described by \cite{maschler2005cold, maschler2008ultracold}
\begin{equation}
\label{eq:dispersiveHamiltonian}
\begin{aligned}
H_1 =& \frac{\hat p_x^2+ \hat p_y^2}{2 m_A} + V_\mathrm{L}^x \cos^2(k_\mathrm{L} \hat x) + V_\mathrm{L}^y \cos^2(k_\mathrm{L} \hat y) \\
&+ V_\mathrm{bias} |\mathcal{F}(\hat x,\hat y)|^2 - \hbar \sum_{m=1}^{M_\mathrm{tot}} \tilde \Delta_{c,m} a_m^\dagger a_m\\
&+ \hbar \sum_{m=1}^{M_\mathrm{tot}}\eta_m \left(h_m^*(\hat x, \hat y) a_m + a_m^\dagger h_m(\hat x, \hat y) \right).
\end{aligned}
\end{equation}
The first line contains the kinetic term with the momentum operators $\hat p_x$ and $\hat p_y$. Classical electric fields create optical potentials with depths $V_\mathrm{L}^x$, $V_\mathrm{L}^y$ and $V_\mathrm{bias}$. The first two create the optical lattice with wave number $k_\mathrm{L}$ and lattice spacing $a = \pi / k_\mathrm{L}$, while $V_\mathrm{bias}$ is much smaller and only responsible for a bias field on certain sites, for instance for excluding diagonals. Thereby $\mathcal{F}(x,y)$ is an electric field distribution whose maximum is normalized to one.

The last two terms describe the free evolution of the cavity fields and atom-state-dependent scattering of the pump fields into the cavity, the atom-light interaction. The quantized electric cavity fields are described by $a_m$ ($a_m^\dagger$), the annihilation (creation) operators of a photon in the $m$-th mode. These fields are coupled to the classical pump fields with mode functions $h_m(x,y)$ via the effective scattering amplitudes $\eta_m = g_m \Omega_m / \Delta_{a,m}$, with the pump laser Rabi frequencies $\Omega_m$, the atom-cavity couplings $g_m$ and the detunings between pump lasers and atomic resonance frequency $\Delta_{a,m}$. The effective cavity detunings $\tilde \Delta_{c,m} = \Delta_{c,m} - N U_{0,m}$ are given by the detunings between pump laser and cavity mode frequencies $\Delta_{c,m}$ and the dispersive shifts of the cavity resonance due to the presence of the atoms in the cavity $N U_{0,m}$ \cite{ritsch2013cold}.

Note that, for example by placing the optical lattice ($x$-$y$-plane) in a common anti-node of the standing wave cavity modes and exciting only TEM$_{00}$ modes, the atom-cavity coupling is uniform in space in our model. Thus the only spatial dependence in the cavity term is due to the pump fields.

\textit{Generalized Bose-Hubbard Hamiltonian.} The atom-atom interactions are taken into account by introducing bosonic field operators $\hat \Psi(\boldsymbol x)$ with $\tilde H = \int \mathrm d^2 x \hat \Psi^\dagger (\boldsymbol x) H_1 \Psi(\boldsymbol x)$ \cite{jaksch1998cold,maschler2005cold}. Note that we do not include contact interactions since atoms never meet due to the initial condition we will use. We assume that the optical lattice with depths $V_\mathrm{L}^x$ and $V_\mathrm{L}^y$ is so deep, that the atoms are tightly bound at the potential minima and only the lowest vibrational state (Bloch band) is occupied. Moreover, the optical potential created by bias and cavity fields is comparably small, such that the form of the Bloch wave functions only depends on the optical lattice \cite{maschler2008ultracold}. In this limit we can expand the bosonic field operators in a localized Wannier basis $\hat \Psi(\boldsymbol x) = \sum_{i,j} w_{2D}(\boldsymbol x - \boldsymbol x_{ij}) b_{ij}$ with the lowest-band Wannier functions $w_{2D}(\boldsymbol x)$ coming from Bloch wave functions of the lattice \cite{kohn1959analytic}. We split the resulting Hamiltonian in three terms
\begin{equation}\label{eq:sweepModes}
\tilde H = H_\mathrm{kin} + H_\mathrm{cav} + H_\mathrm{pot},
\end{equation}
which will be explained in the following.

As in the standard Bose-Hubbard model, one obtains a tunneling term $H_\mathrm{kin}$ as in Eq.\ \eqref{eq:kineticHamiltonian}. Tunneling in $y$-direction is frozen out by ensuring $V_\mathrm{L}^y \gg V_\mathrm{L}^x$. The other terms $H_\mathrm{cav}$ and $H_\mathrm{pot}$ originate from the weak cavity-pump interference fields and the bias fields introduced above and should resemble $H_\mathrm{pr}$ [Eq.\ \eqref{eq:Hpr}]. In order to realize the sweep Eq.\ \eqref{eq:sweep}, the relative strength of these terms and the kinetic term has to be tuned, e.g.\ by ramping up the pump laser and bias field intensity (make $H_\mathrm{cav}$ and $H_\mathrm{pot}$ larger) or the lattice depth (make $H_\mathrm{kin}$ smaller).

The cavity-related terms in Eq.\ (\ref{eq:dispersiveHamiltonian}) give rise to
\begin{equation}
\label{eq:cavhamiltonian}
\begin{aligned}
H_\mathrm{cav} =& -\hbar \sum_m \tilde \Delta_{c,m} a_m^\dagger a_m \\
&+ \hbar \sum_m N \eta_m \left( \hat \Theta_m^\dagger a_m + a_m^\dagger \hat \Theta_m \right)
\end{aligned}
\end{equation}
with the order operator of cavity mode $m$
\begin{equation}
\hat \Theta_m = \frac{1}{N}\sum_{i,j=1}^N \left( v_m^{ij} \hat n_{ij} + u_m^{ij} \hat B_{ij} \right).
\end{equation}
The structure of the fields enters in the on-site and nearest neighbor atom-mode overlaps
\begin{align}\label{eq:modeoverlaps}
v_m^{ij} &= \int dx \, w^2(x-x_i) h_m(x, y_j)\\
u_m^{ij} &= \int dx \, w(x-x_i) h_m(x, y_j) w(x-x_{i+1})
\end{align}
where $y_j = ja$ with $j = 0,...,N-1$ are the tube positions and $w(x)$ the one-dimensional Wannier functions in $x$-direction. This is because for $V_\mathrm{L}^y \gg V_\mathrm{L}^x$ we can approximate the $y$-dependence of the Wannier functions by a Dirac delta: $w_{2D}(\boldsymbol x) = w(x) \delta(y)$.

The last term $H_\mathrm{pot}$ describes all extra fields responsible for local energy off-sets at certain sites that stem from the weak classical fields with the distribution $\mathcal{F}(x,y)$ ($V_\mathrm{bias} \ll V_\mathrm{L}^x$). The off-sets ought to be calculated from the overlaps of fields and Wannier functions, analogously to $v_m^{ij}$. Thus the fields have to be chosen such that the resulting Hamiltonian resembles Eq.\ \eqref{eq:Hpot}. We do not detail the derivation further here and use $H_\mathrm{pot}$ for numerical simulations.

\textit{Atom-atom interaction Hamiltonian.} The main focus of this work is to show how to create the tailored all-to-all particle interactions via collective scattering. We derive this interaction by eliminating the cavity fields introduced in the previous section \cite{nagy2009nonlinear,habibian2013bose,maschler2008ultracold}. This can be done because the cavity fields decay through the mirrors with the rates $\kappa_m$ and thus end up in a particular steady-state for each atomic configuration. Assuming that the atomic motion is much slower than the cavity field dynamics, i.e.\ $J/\hbar \ll |\tilde \Delta_{c,m} + i \kappa_m|$, this steady-state is a good approximation at all times. The stationary cavity field amplitudes are given by
\begin{equation}
a_m^\mathrm{st} \equiv \frac{\eta_m}{\tilde \Delta_{c,m} + i \kappa_m} N \hat \Theta_m
\end{equation}
and thus replaced by atomic operators (see Appendix \ref{app:effectiveHamiltonian}).

In the coherent regime $|\tilde \Delta_{c,m}| \gg \kappa_m$, the atom-light interaction is then described by an effective interaction Hamiltonian for the atoms \cite{caballero2016quantum}
\begin{equation}\label{eq:quantumCavHamiltonian}
H_\mathrm{cav}^\mathrm{eff} = \hbar \sum_m \frac{\tilde \Delta_{c,m} \eta_m^2}{\tilde \Delta_{c,m}^2+\kappa_m^2} N^2 \hat \Theta_m^\dagger \hat \Theta_m.
\end{equation}
The collective, state-dependent scattering induces interactions between each pair of sites $(i,j)$ and $(k,l)$: Density-density interactions due to the terms containing $\hat n_{ij} \hat n_{kl}$ and a modified tunneling amplitude due an occupation or a tunneling event somewhere else in the lattice due to $\hat n_{ij} \hat B_{kl}$, $\hat B_{ij} \hat n_{kl}$ and $\hat B_{ij} \hat B_{kl}$. While the density-density interactions constitute the problem Hamiltonian [see Eq.\ \eqref{eq:HQ}], the latter cavity-induced tunneling terms lead to non-local fluctuations which might help to speed up the annealing process \cite{hormozi2017nonstoquastic}.

Since the Wannier functions are localized at the lattice sites, the on-site overlaps $v_m^{ij}$ tend to be much larger than the nearest-neighbor overlaps $u_m^{ij}$. Thus density-density interactions are expected to be the dominant contribution to $H_\mathrm{cav}^\mathrm{eff}$. As intuitively expected, the atoms localize stronger for deeper lattices, where analytical expressions for the overlaps can be obtained within a harmonic approximation of the potential wells leading to Gaussian Wannier functions with a width $\propto (V_\mathrm{L}^x)^{-1/4}$. Apart from a correction factor due to this width, the on-site overlaps are given by the pump fields at the lattice sites. The nearest-neighbor overlaps correspond to the pump fields in between the lattice sites, but are exponentially suppressed (see Appendix \ref{app:deeplatticelimit}). Consequently, in the deep lattice limit (large $V_\mathrm{L}^x$) when the width tends to zero we get $v_m^{ij} = h_m(x_i,y_j)$ and $u_m^{ij} = 0$ \cite{caballero2016quantum}, and an interaction Hamiltonian [from Eq.\ \eqref{eq:quantumCavHamiltonian}]
\begin{equation}\label{eq:quantumCavHamiltonianDL}
\begin{aligned}
H_\mathrm{cav}^\mathrm{dl} = \hbar \sum_m& \frac{\tilde \Delta_{c,m} \eta_m^2}{\tilde \Delta_{c,m}^2+\kappa_m^2} \\
\times &\sum_{ijkl} h_m^*(x_i,y_j) h_m(x_k, y_l) \hat n_{ij} \hat n_{kl},
\end{aligned}
\end{equation}
which only depends on density operators, and hence does not include cavity induced-tunneling.

\subsection{$N$-queens interaction}
\label{subsec:nqueensinteraction}

In this section we aim to find pump fields $h_m(x,y)$ such that the interaction Hamiltonian in the deep lattice limit Eq.\ \eqref{eq:quantumCavHamiltonianDL} corresponds to the desired queens Hamiltonian $H_\mathrm{Q}$ [Eq.\ (\ref{eq:HQ})] containing the non-attacking conditions. Using these pump fields we later show numerically in Sec.\ \ref{sec:justification}, that the atom-atom interaction for realistic lattice depths [Eq.\ \eqref{eq:quantumCavHamiltonian}], although slightly altered, still well resembles the queens interaction.

We consider three sets of $M$ parallel running wave laser beams with different propagation directions, each of which could be created by a frequency comb. The three directions are perpendicular to the lines along which queens should not align, that is along the $x$-direction and along the diagonals. We denote the corresponding wave vectors with $\boldsymbol k_m^x = (k_m^0, 0)^T$, $\boldsymbol k_m^+ = (k_m^0, k_m^0)^T$ and $\boldsymbol k_m^- = (k_m^0, -k_m^0)^T$, respectively, with the wave numbers $k_m^0$. Therefore the pump fields are given by
\begin{equation}\label{eq:runningwaves}
h_m(x,y) = e^{i \boldsymbol{k}_m \boldsymbol{x}},
\end{equation}
where $\boldsymbol x = (x,y)^T$ is the position vector and $\boldsymbol k_m$ is a wave vector in any of the three directions.

With running wave pump fields, Eq. \eqref{eq:quantumCavHamiltonianDL} can be written as
\begin{equation}\label{eq:Hdd}
H_\mathrm{cav}^\mathrm{dl} = U_\mathrm{Q} \sum_{ijkl} \tilde A_{ijkl} \hat n_{ij} \hat n_{kl}.
\end{equation}
This formally corresponds to $H_\mathrm{Q}$ [Eq.\ (\ref{eq:HQ})], where the quantities now have a physical meaning: The interaction matrix is given by
\begin{equation}\label{eq:tildeA}
\tilde A_{ijkl} = \sum_m f_m \cos(\boldsymbol k_m (\boldsymbol x_{ij} - \boldsymbol x_{kl}))
\end{equation}
with lattice site connection vectors $\boldsymbol x_{ij} - \boldsymbol x_{kl}$ and
\begin{equation}
f_m U_\mathrm{Q} = \hbar \frac{\tilde \Delta_{c,m} \eta_m^2}{\tilde \Delta_{c,m}^2+\kappa_m^2}
\end{equation}
with $\sum_{m=0}^{M-1} f_m = 1$. The dimensionless parameters $f_m$ capture the relative strengths of the modes, determining the shape of the interaction. They have to be chosen such that $\tilde A$ approximates $A$ [Eq.\ \eqref{eq:A}]. The overall strength of the interaction term is captured by the energy $U_\mathrm{Q}$, which can be easily tuned by the cavity detunings or the pump intensities to implement the parameter sweep in Eq.\ \eqref{eq:sweep}. For the following discussion we define an interaction function
\begin{equation}\label{eq:intfunc}
\tilde{\mathcal{A}} (\boldsymbol r) = \sum_m f_m \cos(\boldsymbol k_m \boldsymbol r)
\end{equation}
which returns the interaction matrix when evaluated at lattice site connection vectors $\tilde A_{ijkl} = \tilde{\mathcal{A}} (\boldsymbol x_{ij} - \boldsymbol x_{kl})$.

We note that one set of parallel $\boldsymbol k_m^\mu$ ($\mu \in \{x,+,-\}$) creates an interaction $\tilde{\mathcal{A}}$ which is constant and infinite range (only limited by the laser beam waist) in the direction perpendicular to the propagation direction $\boldsymbol r \perp \boldsymbol k_m^\mu$. Along the propagation direction $\boldsymbol r \parallel \boldsymbol k_m^\mu$ instead, the interaction is shaped according to the sum of cosines, and can be modified by the choice of wave numbers $k_m^\mu = |\boldsymbol k_m^\mu|$ and their relative strengths $f_m$.

\begin{figure}[tb]
  \centering
  \includegraphics[width=\columnwidth]{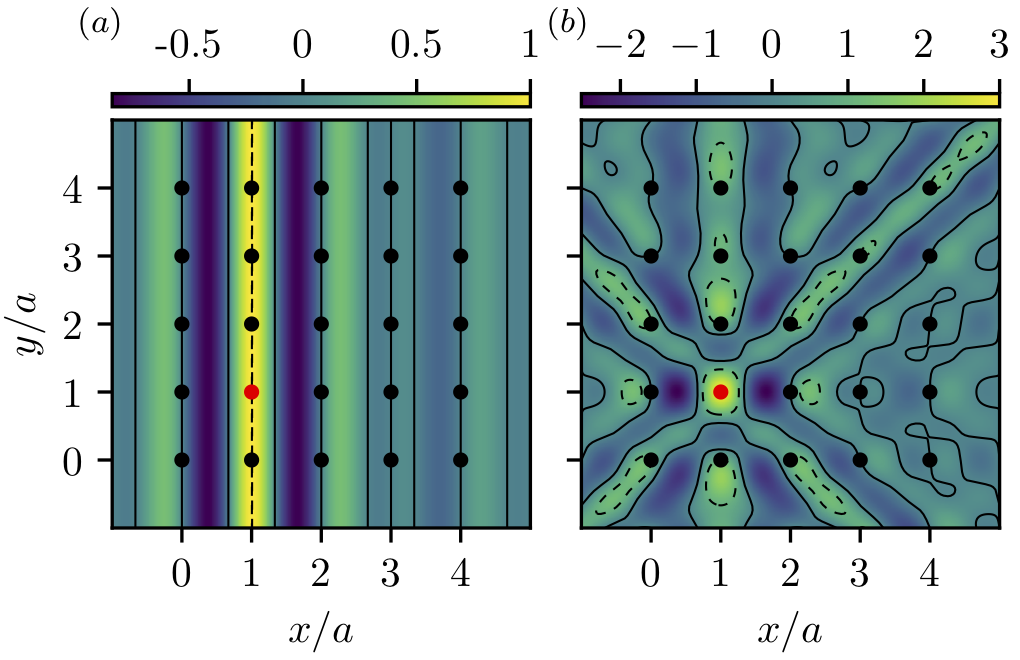}
  \caption{{\it Energy penalty created by one atom.} - These density plots show the energy penalty $\tilde{\mathcal{A}}(\boldsymbol x - \boldsymbol x_a)$ for an atom at position $\boldsymbol x = (x,y)^T$ created by an atom at position $\boldsymbol x_a = (a,a)^T$ (red dot) for $N=M=5$ [see Eq.\ (\ref{eq:intfunc})]. The dots indicate lattice site positions and the solid (dashed) contour lines indicate where $\tilde{\mathcal{A}}(\boldsymbol x - \boldsymbol x_a) = 0$ ($\tilde{\mathcal{A}}(\boldsymbol x - \boldsymbol x_a) = 1$) is fulfilled. (a) Pumping along the $x$-axis with the wave vectors $\boldsymbol k_m^x = (k_m^0, 0)^T$ with $k_m^0/k_\mathrm{L} = 1.1, 1.3, 1.5, 1.7, 1.9$ [according to Eq.\ \eqref{eq:comb}] creates interactions along $y$. (b) Additionally including diagonal pump lasers $\boldsymbol k_m^+ = (k_m^0, k_m^0)^T$ and $\boldsymbol k_m^- = (k_m^0, -k_m^0)^T$ implements the full queens interaction along diagonals and vertical lines [Eq.\ \eqref{eq:HQ}].}
  \label{fig:interactions}
\end{figure}

In the following we will use the example wave numbers
\begin{equation}\label{eq:comb}
k_m^0 = k_\mathrm{L} \left(1 + \frac{2m+1}{2M}\right)
\end{equation}
with $m = 0,...,M-1$ and uniform $f_m = 1/M$. Taking into account $\boldsymbol k_m^x$ only, the interaction along the $x$-direction $\boldsymbol r \parallel \boldsymbol k_m^x$ at lattice site distances $r_j = j \pi/k_\mathrm{L} = ja$ has the values
\begin{equation}\label{eq:combproperty}
\tilde{\mathcal{A}}(r_j) = \begin{cases}
(-1)^l \; &\text{for} \; j = 2 M l, \; l \in \mathbb{Z}\\
0 \; &\text{otherwise},
\end{cases}
\end{equation}
as shown in Appendix \ref{app:interaction}. If we guarantee that $-2M<j<2M$ this results in an interaction which is zero everywhere apart from $j = 0$, i.e.\ at zero distance. So for repulsive interactions ($U_\mathrm{Q} > 0$ and thus $\tilde \Delta_{c,m} > 0$), the wave vectors $\boldsymbol k_m^x$ create the non-attacking interaction along the $y$-direction ($\tilde A_{ijkl} = 1$ if $i=k$ and $0$ otherwise) as long as $N \leq 2M$. This is illustrated in Fig.\ \ref{fig:interactions}(a) for $M=N=5$. Analogously, $\boldsymbol k_m^\pm$ cause the non-attacking interactions along the diagonals. In a square lattice the diagonals have the distance $r_j/\sqrt{2}$, which is compensated by $k_m^\pm = |\boldsymbol k_m^\pm| = \sqrt{2} k_m^0$. Since there are $2N-1$ diagonals, one has to make sure that $2N-1 \leq 2M$. Upon combining all wave vectors from three directions we finally obtain the full queens interaction, as shown in Fig.\ \ref{fig:interactions}(b), which is realized with $M_\mathrm{tot} = 3M = 3N$ frequencies in our example.

Note that there are several combinations of wave numbers and mode strengths which, at least approximately, create the desired line-shaped interactions perpendicular to the light propagation direction. For this it is insightful to reformulate the interaction as a Fourier transform. To deal with continuous functions, we define an envelope $f(k)$ with $f(k_m) = f_m$, which is sampled at the wave numbers $n\Delta k$ with $n \in \mathbb{Z}$ containing all $k_m^\mu$. Considering one illumination direction for simplicity, the interaction [Eq.\ (\ref{eq:intfunc})] along $\boldsymbol r \parallel \boldsymbol k_m^\mu$ with $r = |\boldsymbol r|$ can be written as
\begin{equation}
\begin{aligned}
\tilde{\mathcal{A}}(r) &= \mathrm{Re} \left[ \sqrt{2\pi} \mathcal{F} \left\{ f(k) \sum_{n=-\infty}^\infty \delta(k - n \Delta k) \right\}(r) \right]\\
&= \sum_{l=-\infty}^\infty \mathrm{Re}\left[\sqrt{2\pi} \mathcal{F}\{f\}\left(r - l \frac{2\pi}{\Delta k}\right)\right],
\end{aligned}
\end{equation}
where $\mathcal{F}\{f\}(r) = \int_{-\infty}^\infty \mathrm dk f(k) e^{ikr}/\sqrt{2\pi}$ is the Fourier transform of $f(k)$ and $\delta(x)$ is a Dirac delta at $x=0$. See Appendix \ref{app:interaction} for a detailed derivation.

The last line allows for a simple interpretation: The interaction consists of peaks repeating with a spatial period $R = 2\pi/\Delta k$. Each of these peaks has the shape of the real part of the Fourier transform of the envelope function $f(k)$ with a width corresponding to the inverse of the mode bandwidth $\sigma \sim 2\pi/\Delta k_\mathrm{BW}$.

For the (approximate) non-attacking condition ($\tilde{\mathcal{A}}(ja) \approx 1$ for $j=0$ and $|\tilde{\mathcal{A}}(ja)| \ll 1$ otherwise) there are two conditions. Firstly, at most one peak should be within the region of the atoms. Thus the period has to be larger than the (diagonal) size of the optical lattice $R \geq Na$ ($R \geq \sqrt{2}Na$). Secondly, the width of one peak has to be smaller than the lattice spacing $\sigma \lesssim a$. Combining these conditions to $R \gtrsim N\sigma$, we see that the minimum number of modes per direction $M$ scales linearly with $N$
\begin{equation}
M \approx \Delta k_\mathrm{BW} / \Delta k \gtrsim N.
\end{equation}
Therefore, with only $\sim N$ modes this quite generically allows for creating an interaction along lines perpendicular to the light propagation. Note however, that the second condition also implies that the spatial frequency spread has to be at least on the order of the lattice wave number $\Delta k_\mathrm{BW} \gtrsim k_\mathrm{L}$.

\section{Numerical justification of assumptions}
\label{sec:justification}
We compare the ideal model Hamiltonian described in Sec.\ \ref{sec:model} [Eq.\ \eqref{eq:sweep}] to the physically motivated tight-binding Hamiltonian for finite lattice depths introduced in Sec.\ \ref{sec:implementation} [Eq.\ \eqref{eq:sweepModes}]. As for the ideal model in Fig.\ \ref{fig:timeevolution}, we consider the time evolution during a slow linear sweep of $U_\mathrm{Q}$, $U_\mathrm{T}$ and $U_\mathrm{D}$ by numerically integrating the time-dependent Schr\"odinger equation. Physically, this sweep can be realized by ramping up the pump and the bias field intensities. Moreover, we show that evolving the system using a classical approximation for the cavity mode fields does not result in a solution to the $N$-queens problem, and address the effect of dephasing by photon loss with open system simulations.

In the following we use a realistic lattice depth of $V_\mathrm{L}^x = 10 E_\mathrm{R}$ with the recoil energy $E_\mathrm{R} = \hbar^2 k_\mathrm{L}^2/(2m_\mathrm{A})$. For example, for rubidium $^{87}$Rb and $\lambda_\mathrm{L} = 785.3 \, \mathrm{nm}$ it is $E_\mathrm{R}/\hbar = 23.4 \, \mathrm{kHz}$ \cite{landig2016quantum}. The chosen lattice depth leads to a tunneling amplitude $J \approx 0.02 E_\mathrm{R}$, which can be obtained from the band structure of the lattice. We consider our cavity model in Eq.\ \eqref{eq:sweepModes} for $N=5$. The pump modes are as in Sec.\ \ref{subsec:nqueensinteraction} and Fig.\ \ref{fig:interactions}. While in the limit of a deep lattice this would result in the ideal model interactions, here they depend on the overlaps between Wannier functions and pump modes [Eq.\ \eqref{eq:modeoverlaps}] and are thus altered. In the following the overlaps are calculated with Wannier functions which where numerically obtained from the band structure of the lattice. It turns out, that the deviation from the ideal overlaps does not qualitatively change the interaction for the realistic parameters used.

\subsection{Coherent dynamics}
\label{ssec:timeevo_sch}
\begin{figure}[tb]
  \centering
  \includegraphics[width=\columnwidth]{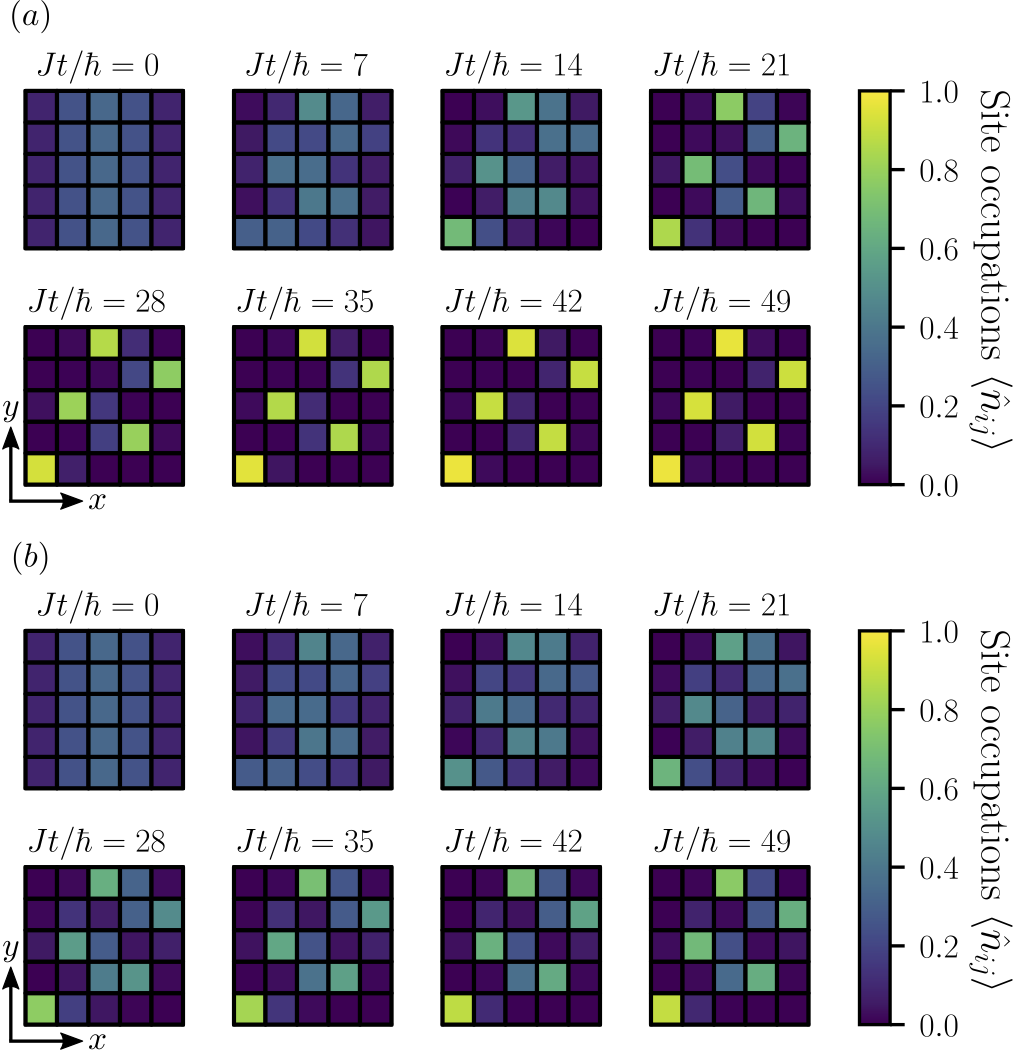}
  \caption{\textit{Comparison of the time evolution with quantum and classical fields.} - Lattice site occupations for a time evolution during a nearly adiabatic sweep using the cavity Hamiltonian including cavity-assisted tunneling for the same parameters as in Fig.\ \ref{fig:timeevolution}. Subplot (a) shows the dynamics using the full quantum interaction Hamiltonian [Eq.\ \eqref{eq:quantumCavHamiltonian}]. It closely resembles the results from the model Hamiltonian in Fig.\ \ref{fig:timeevolution}. Subplot (b) shows the time evolution with the classical approximation of the cavity fields [Eq.\ \eqref{eq:classicalCavHamiltonian}]. The state does not converge to the solution, also not for much larger sweep times. The modes for both cases where chosen as in Fig.\ \ref{fig:interactions}(b).}
  \label{fig:timeevolution_comparison}
\end{figure}

The energy spectrum of the Hamiltonian for finite lattice depths in Eq.\ \eqref{eq:quantumCavHamiltonian} is shown in Fig.\ \ref{fig:spectrum}(b) and is qualitatively of the same form as for the ideal model in Fig.\ \ref{fig:spectrum}(a). In comparison the eigenvalue gaps tend to be smaller at the end of the sweep. This is because the on-site atom-mode overlaps decrease for shallower lattices and less localized atoms due to a smoothing of the mode functions by the finite width Wannier functions (see Appendix \ref{app:deeplatticelimit}). Moreover, we consider the time evolution during the nearly adiabatic sweep for $J\tau / \hbar= 49$ for the same parameters by integrating the time-dependent Schr\"odinger equation. Snapshots of the site occupations $\langle \hat n_{ij} \rangle$ for several times are shown in Fig.\ \ref{fig:timeevolution_comparison}(a), where we observe a similar behavior as for the model in Fig.\ \ref{fig:timeevolution}. This suggests that the system is robust against the errors introduced by the moderate lattice depth. Note that the physical Hamiltonian used here is non-stoquastic in the occupation basis due to cavity-induced tunneling.

\subsection{Classical cavity fields}
\label{ssec:timeevo_class}
In the following we show that quantum correlations are crucial for the efficiency of the sweep. In particular, if we substitute the field operators $a_m^\mathrm{st}$ by its expectation values representing classical cavity fields the solution is not found. We consider the semi-classical Hamiltonian
\begin{equation}
\label{eq:classicalCavHamiltonian}
\begin{aligned}
H_\mathrm{cav}^\mathrm{class} =& \hbar \sum_m \frac{N^2 \tilde \Delta_{c,m} \eta_m^2}{\tilde \Delta_{c,m}^2+\kappa_m^2} \\
& \times \left(\hat \Theta_m^\dagger \langle \hat \Theta_m \rangle + \langle \hat \Theta_m^\dagger \rangle \hat \Theta_m - |\langle \hat \Theta_m \rangle|^2 \right),
\end{aligned}
\end{equation}
where the expectation values have to be calculated self-consistently with the current atom state vector. This substitution amounts to considering only first order fluctuations around the mean of $\hat \Theta_m$ in Eq.\ (\ref{eq:quantumCavHamiltonian}).

Consequently, the dynamics are described by a differential equation which is non-linear in the state vector $|\psi\rangle$. We numerically solve this equation by self-consistently updating the expectation value in each time step. It turns out that even for very long sweep times, using classical fields does not lead to a solution of the queens problem. The time evolution for $J\tau / \hbar = 49$ is depicted in Fig.\ \ref{fig:timeevolution_comparison}(b). The discrepancy shows the necessity of entangled light-matter states in our procedure.

\subsection{Dephasing due to cavity field loss}
\label{ssec:timeevo_open}
Finally, motivated by experimental considerations, we consider the open system including photon decay through the cavity mirrors. This system can be described by a master equation for the atoms \cite{maschler2008ultracold}
\begin{equation}\label{eq:master}
\begin{aligned}
\dot \rho =& -\frac{i}{\hbar}[H_\mathrm{kin} + H_\mathrm{cav}^\mathrm{eff},\rho] \\
			&+ \sum_m  \frac{N^2 \eta_m^2\kappa_m}{\Delta_{c,m}^2+\kappa_m^2} \left( 2 \hat \Theta_m \rho \hat \Theta_m^\dagger - \{ \hat \Theta_m^\dagger \hat \Theta_m, \rho \} \right),
\end{aligned}
\end{equation}
where curly brackets denote the anti-commutator (see Appendix \ref{app:effectiveHamiltonian}). The model is suitable for analyzing the dephasing close to the coherent regime before any steady state is reached.

More insight can be gained by rewriting it in the basis of scattering eigenstates $|\nu\rangle$ with $\hat \Theta_m |\nu\rangle = \theta_m^\nu |\nu\rangle$, which scatter a field $\alpha_m^\nu = \frac{N \eta_m}{\tilde \Delta_{c,m}+i\kappa_m} \theta_m^\nu$. These states converge to the occupation states in the deep lattice limit. The time evolution for the matrix elements $\rho_{\mu\nu} = \langle \mu | \rho | \nu \rangle$ reads
\begin{equation}
\begin{aligned}
\dot \rho_{\mu\nu} =& (- \Gamma_{\mu\nu} - i \Omega_{\mu\nu}) \rho_{\mu\nu} \\
&- i \frac{J}{\hbar} \sum_k \left(\langle \mu |  \hat B | k \rangle \rho_{k\nu} - \langle k | \hat B | \nu \rangle \rho_{\mu k}\right)
\end{aligned}
\end{equation}
with the rates
\begin{align}
\Omega_{\mu\nu} =& \sum_m \tilde \Delta_{c,m} (|\alpha_m^\mu|^2-|\alpha_m^\nu|^2)\\
\Gamma_{\mu\nu} =& \sum_m \kappa_m |\alpha_m^\mu - \alpha_m^\nu|^2.
\end{align}
While the energy gaps $\hbar \Omega_{\mu\nu}$ describe the coherent dynamics we considered up to now, the dephasing rates $\Gamma_{\mu\nu}$ stem from photon loss. Hence a superposition of two scattering eigenstates $|\mu\rangle$ and $|\nu\rangle$ looses its coherence depending on the difference of the scattered fields, or how distinguishable the states are by field measurement.

\begin{figure}[tb]
  \centering
  \includegraphics[width=\columnwidth]{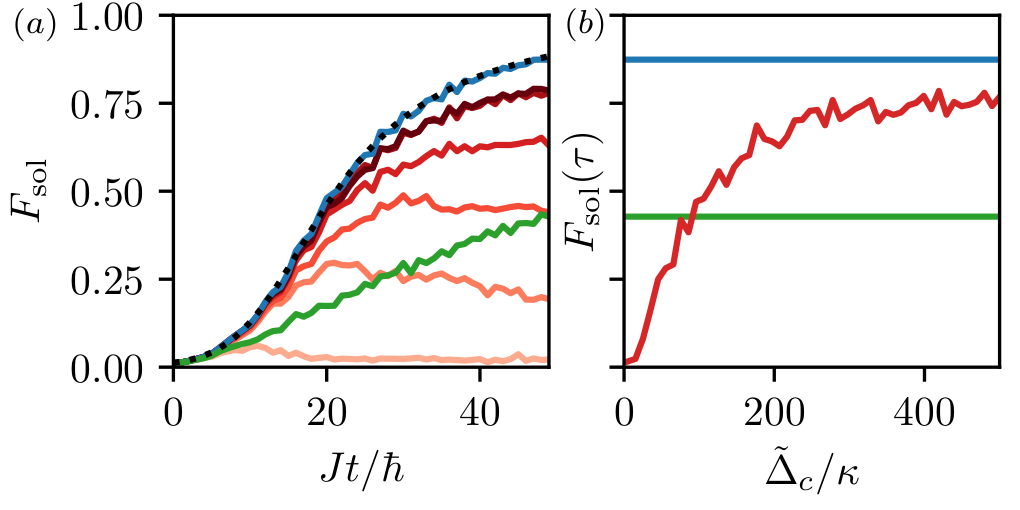}
  \caption{\textit{Comparison to open system dynamics.} - (a) The figure shows the fidelity $F_\mathrm{sol}$ between the solution and the instantaneous eigenstate (black dotted), and between solution and dynamical state from the time evolution using the Schr\"odinger equation (blue), classical fields (green) and the open system (red). The open system fidelity depends on $\tilde \Delta_c/\kappa$, where we present results for the values $5,50,100,200,500,1000$ (from light red to dark red). (b) The fidelity between solution and the final state after the sweep as a function of $\tilde \Delta_c/\kappa$ for the curves of (a), where the colors are the same as in (a). The parameters used here are as in Fig.\ \ref{fig:timeevolution_comparison}.}
  \label{fig:fidelity}
\end{figure}

We now return to the example we had before, using uniform mode strengths $f_m = 1/M$, decay rates $\kappa$ and detunings $\tilde \Delta_{c}$. Figure \ref{fig:fidelity} compares the coherent Schr\"odinger time evolution from Sec.\ \ref{ssec:timeevo_sch}, the mean-field approximation with classical cavity fields from Sec.\ \ref{ssec:timeevo_class} and the open system dynamics with dephasing. For the latter the master equation is approximated by using Monte-Carlo wave function simulations. As a measure of similarity between states we use the fidelity. For two mixed states it is defined as $F(\rho,\sigma) = \mathrm{Tr}(\sqrt{\sqrt{\rho} \sigma \sqrt{\rho}})$ and reduces to the overlap $|\langle \psi | \phi \rangle|$ for pure states.

The open system dynamics is depicted for different detunings $\tilde \Delta_c / \kappa$ while keeping $U_\mathrm{Q}$ fixed. This can be achieved by adjusting the pump strength $\eta$ correspondingly. In this case the dephasing rates
\begin{equation}
\frac{\hbar \Gamma_{\mu\nu}}{U_\mathrm{Q}} = \frac{\kappa}{\tilde \Delta_{c}} N^2 \sum_m |\theta_m^\mu - \theta_m^\nu|^2
\end{equation}
go to zero for $\tilde \Delta_c / \kappa \gg 1$. Thus, as expected, the open system converges to the coherent Schr\"odinger dynamics in this limit (see Fig.\ \ref{fig:fidelity}(b)).

Note that the coherence between states creating similar fields is preserved much longer than for other states, which is expected to be important at the late stage of the sweep. For states with fixed similarity (e.g.\ one atom moved), $|\theta_m^\mu - \theta_m^\nu|^2$ is on the order of $N^{-2}$, and thus the dephasing rates do not scale with $N$ for such states.

\section{Read-out}
\label{sec:readout}
After the parameter sweep we need to determine if the obtained state is a solution or not. This can in principle be done by reading out the final atomic state with single site resolution using a quantum gas microscope \cite{bakr2009quantum,sherson2010single}. However, as we consider an open system with the cavity output fields readily available, we will show that by proper measurements on the output light we can directly answer this question without further additions. Note that after the sweep at the stage of the read-out, quantum coherences do not have to be preserved since the solution is a classical state. This gives the freedom to increase the lattice depth to some high value in the deep lattice regime to suppress further tunneling, and to increase the pump power or decrease the detunings in order to get a stronger signal at the detector.

\subsection{Intensity measurement}
For uniform cavity detunings, a state corresponding to the solution of the $N$-queens problem scatters less photons than all other states. Thus the measurement of the total intensity in principle allows one to distinguish a solution from other states. To illustrate this we consider the total rate of photons impinging on a detector scattered by an atomic state $|\psi \rangle$
\begin{equation}
\begin{aligned}
P(|\psi\rangle) =& \sum_m 2 \kappa_m \langle (a_m^\mathrm{st})^\dagger a_m^\mathrm{st} \rangle \\
=& \sum_m \frac{2\kappa_m}{\tilde \Delta_{c,m}} \frac{\tilde \Delta_{c,m}\eta_m^2 N^2}{\tilde \Delta_{c,m}^2 + \kappa_m^2} \langle \hat \Theta_m^\dagger \hat \Theta_m \rangle \\
\approx & U_\mathrm{Q} \frac{\zeta}{\hbar} \sum_{ijkl} \tilde A_{ijkl} \langle \hat n_{ij} \hat n_{kl} \rangle = \frac{\zeta}{\hbar} \langle H_\mathrm{cav}^\mathrm{dl} \rangle.
\end{aligned}
\end{equation}
In the last line we assumed that $\zeta = 2\kappa_m / \tilde \Delta_{c,m}$ does not depend on $m$ and a deep lattice.

Since $P$ is proportional to the energy expectation value, the ground state, i.e.\ the solution of the queens problem, causes a minimal photon flux at the detector $P_0 = 3N U_\mathrm{Q} \zeta$. It stems from the on-site terms $(i,j)=(k,l)$, where the factor $3$ comes from the three pump directions. In contrast, each pair of queens violating the non-attacking condition in $\tilde A$ leads to an increase of the photon flux by $\Delta P = 2 U_\mathrm{Q} \zeta$. The two atoms create an energy penalty for one another, explaining the factor $2$. The relative difference of the photon flux due to a state with $L$ attacking pairs and a solution is given by
\begin{equation}
\frac{L\Delta P}{P_0} = \frac{2L}{3N}.
\end{equation}
As this scales with $1/N$ it is difficult to distinguish solutions from other states via measurement of the intensity for large $N$. Note that for non-uniform $\kappa_m / \tilde \Delta_{c,m}$, photons from different modes have to be distinguished.

\begin{figure}[h!]
  \centering
  \includegraphics[width=\columnwidth]{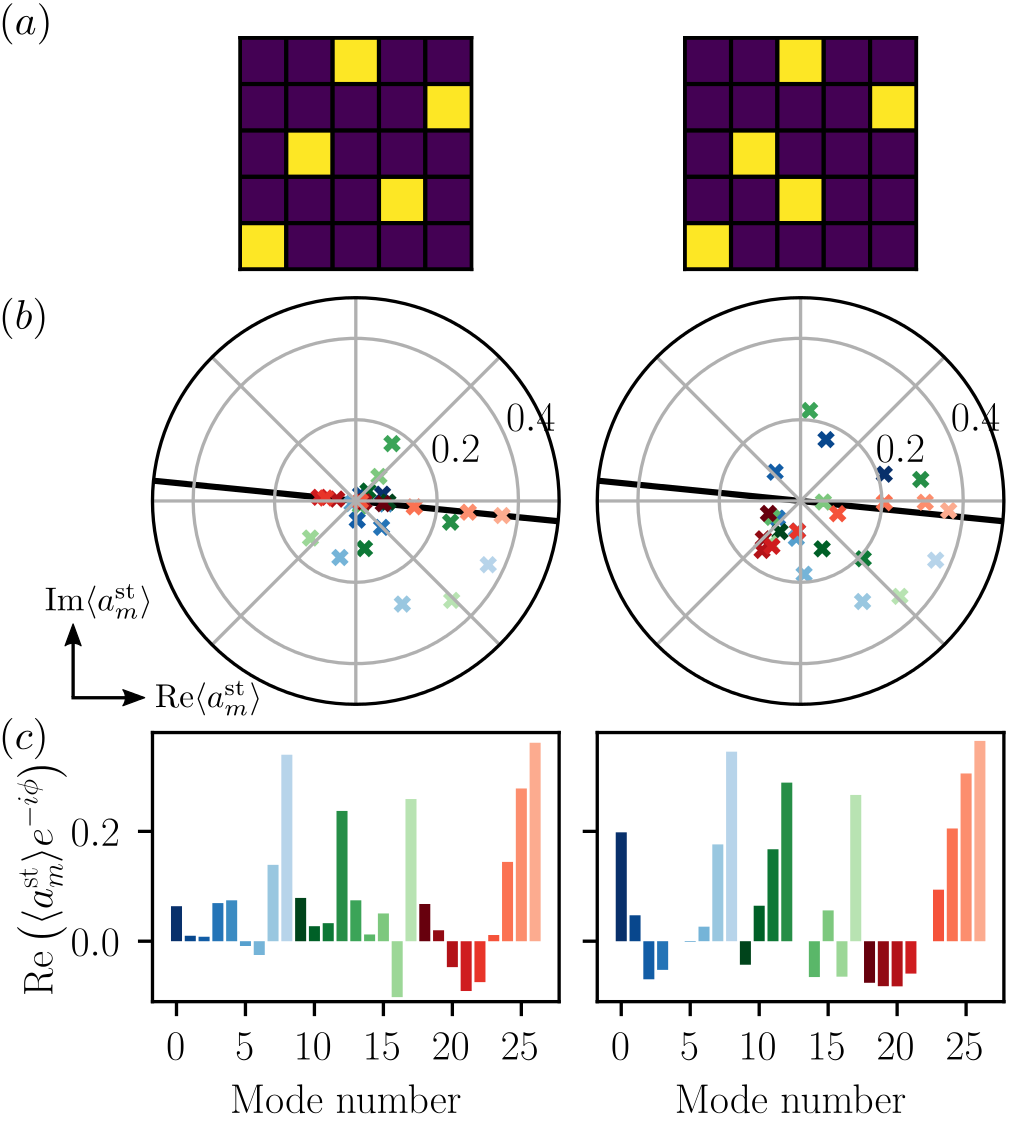}
\caption{\textit{Signature of an atomic state in the cavity output.} - In left and right column the signature of two different atomic (pure) states are compared, whose occupations are shown on the top (a). On the left there is a solution to the $N$-queens problem, while on the right one atom was moved. (b) The polar plot shows the cavity field expectation value $\langle a_m^\mathrm{st} \rangle$ in the complex plane. Measuring the fields by homodyne detection yields a certain quadrature of the field depending on the phase angle $\phi$. An example is illustrated by the black line. (c) The measurable quadratures are shown for the example angle. The sequences can be Fourier transformed to obtain the occupations of columns and of the diagonals. The parameters used here are $U_\mathrm{Q} = 5J$ and $\tilde \Delta_{c}/\kappa = 10$. We used $M = 2N-1 = 9$ modes per direction with wave numbers as in Eq.\ \eqref{eq:comb}. Different colors encode the different directions of the pump modes: $\boldsymbol k_m^x$ (blue), $\boldsymbol k_m^+$ (green) and $\boldsymbol k_m^-$ (red). The lighter the color the larger the wave number $|\boldsymbol k_m^{x,\pm}|$.}
  \label{fig:fields}
\end{figure}

\subsection{Field measurement}

Measurement of the output field quadratures, for example by homodyne detection, gives insight about the absolute position of the atoms projected onto the pump laser propagation direction. For the three directions used in our setup, this yields the occupations of each column $\langle \hat N_i^x \rangle$ and each diagonal $\langle \hat N_i^+ \rangle$ and $\langle \hat N_i^- \rangle$. Since a solution of the queens problem has maximally one atom on each diagonal and exactly one atom on each column, it must fulfill
\begin{equation}\label{eq:condition}
\langle \hat N_i^x \rangle = 1 \land \langle \hat N_i^+ \rangle \leq 1 \land \langle \hat N_i^- \rangle \leq 1.
\end{equation}
The output field quadratures can thus be used to determine if a classical final state is a solution or not, which is the answer to the blocked diagonals decision problem we aim to solve. The signatures of two example states in the cavity fields are depicted in Fig.\ \ref{fig:fields}.

Let us illustrate the measurement by considering only light scattered from the $x$-direction with incident wave vectors $\boldsymbol k_m^x$. Since these plane waves are constant in $y$-direction, the atom-field overlaps do not depend on $j$. Neglecting cavity-induced tunneling, the field quadratures for a phase difference $\phi$ are
\begin{equation}\label{eq:field_measurement}
\begin{aligned}
\mathrm{Re}(\langle a_m^\mathrm{st} \rangle e^{-i\phi}) &= \sum_i \mathrm{Re} \left( \frac{\eta_m e^{-i\phi}}{\tilde \Delta_{c,m} + i \kappa_m} v_m^{i1} \right) \langle \hat N_i^x \rangle,
\end{aligned}
\end{equation}
revealing that cavity fields are determined by the total occupations of the columns $\hat N_i^x = \sum_j \hat n_{ij}$.

For at least $N$ modes ($M \geq N$) this system of equations can be inverted yielding the column occupations $\langle \hat N_i^x \rangle$. By measuring the cavity output field quadratures scattered from the diagonal pump light we obtain the occupations of each diagonal $\langle \hat N_i^+ \rangle$ and $\langle \hat N_i^- \rangle$. Inverting the system of equations for diagonals demands at least as many pump modes as diagonals, that is $2N-1$. Inversion can also be done efficiently and intuitively by using a discrete Fourier transform and its inverse. To reveal the Fourier relation, one has to express the $v_m^{ij}$'s in Eq.\ \eqref{eq:field_measurement} within the harmonic (or in the deep lattice) approximation (see Appendix \ref{app:deeplatticelimit}). The so obtained approximate inversion formula also works well for realistic lattice depths.

Strictly speaking the condition in Eq.\ \eqref{eq:condition} is sufficient only for classical configurations, like occupation number basis states $|\phi_\nu\rangle$. Some superpositions $|\psi\rangle = \sum_\nu c_\nu |\phi_\nu\rangle$ which are no solutions might also fulfill the above criterion, because summands in the field expectation values $\langle a_m^\mathrm{st} \rangle = \sum_\nu |c_\nu|^2 \langle \phi_\nu | a_m^\mathrm{st} | \phi_\nu \rangle$ can cancel each other. For instance, for $U_\mathrm{T}=0$ the solution from our example in Fig.\ \ref{fig:timeevolution} $|\psi_\mathrm{sol}\rangle = |1,4,2,5,3\rangle$ scatters the same fields $\langle a_m^\mathrm{st} \rangle$ as the superposition $|\psi_\mathrm{nosol}\rangle = (|\psi^1_\mathrm{nosol} \rangle + |\psi^2_\mathrm{nosol} \rangle)/\sqrt{2}$ with $|\psi^1_\mathrm{nosol} \rangle = |1,3,2,5,4\rangle$ and $|\psi^2_\mathrm{nosol} \rangle = |1,4,5,2,3\rangle$, both of which are no solution. In this notation the state $|i_1, i_2, ..., i_N\rangle$ has one atom on each site $(i_j,j)$. However, these macroscopic superpositions are highly unstable. Even theoretically the measurement back-action \cite{carmichael1993open,mekhov2009quantum} projects superpositions of states scattering different fields (such as $|\psi_\mathrm{nosol}\rangle$) to one of its constituents. The inclusion of measurement back-action due to continuous measurement might thus lead to intriguing phenomena beyond those presented here and is subject to future work.

We emphasize again that the measurements described above answer the question if we found a solution or not, which is the answer to the combinatorial decision problem. The exact configuration of the final state can be measured with single site resolution as demonstrated in several experiments \cite{bakr2009quantum,sherson2010single}.

\section{Conclusions}
\label{sec:conclusions}
We present a special purpose quantum simulator with the aim to solve variations of the $N$-queens problem based on atoms in a cavity. This combinatorial problem may serve as a benchmark to study a possible quantum advantage in intermediate size near term quantum experiments. From the algorithmic point of view, the problem is interesting for quantum advantage as it is proven NP-hard and instances can be found that are not solvable with current state-of-the-art algorithms. From the implementation point of view, the proposed quantum simulator implements the queens problem without overhead and thus a few tens of atoms are sufficient to enter the classically intractable regime. The proposed setup of atoms in a cavity fits the queens problem naturally as the required infinite range interactions arise there inherently. We find that by treating the light field classically the simulation does not find the solutions suggesting that quantum effects like atom-field entanglement cannot be neglected. Moreover, we investigate the influence of photon loss on the coherence time.

The queens problem is formulated as a decision problem, asking whether there is a valid configuration of queens or not given the excluded diagonals and fixed queens. Remarkably, to answer the decision problem, a read-out of the atom positions is not required as the necessary information is encoded in the light that leaves the cavity. To determine the position of the queens requires single site resolved read-out, which is also available in several current experimental setups \cite{bakr2009quantum}.

In this work we concentrated on the coherent regime. The driven-dissipative nature of the system provides additional features which can be exploited for obtaining the ground state. For certain regimes, cavity cooling \cite{ritsch2013cold,wolke2012cavity} can help to further reduce sweep times and implement error correction. Moreover, the back action of the field measurement onto the atomic state can be used for preparing states \cite{mekhov2009quantum}.

Note that an implementation of the $N$-queens problem for a gate-based quantum computer was proposed in Ref.\ \cite{jha2018novel} aiming to find a solution of the unconstrained $N$-queens problem. Our work in contrast employs an adiabatic protocol and intends to answer the question if a solution exists given constraints of blocked diagonals or already placed queens, which was shown to be NP-complete and numerically hard \cite{gent2017complexity}.

\textit{Acknowledgments.} We thank I.\ Gent, C.\ Jefferson and P.\ Nightingale for fruitful discussions. Simulations were performed using the open source QuantumOptics.jl framework in Julia \cite{kramer2018toolbox} and we thank D.\ Plankensteiner for related discussions. V.\ T.\ and H.\ R.\ are supported by Austrian Science Fund Project No.\ I1697-N27. W.\ L.\ acknowledges funding by the Austrian Science Fund (FWF) through a START grant under Project No.\ Y1067-N27 and the SFB BeyondC Project No.\ F7108-N38, the Hauser-Raspe foundation, and the European Union's Horizon 2020 research and innovation program under grant agreement No.~817482 PasQuanS.


\appendix
\onecolumngrid

\newpage

\section*{Appendix}

\section{Instance parameters} \label{app:parameters}
Table \ref{tbl:Values} provides an overview of the chosen parameters for the exemplary linear parameter sweep in the main text (Figs. \ref{fig:timeevolution}, \ref{fig:spectrum} and \ref{fig:timeevolution_comparison}).
\begin{table}[htb]
	\centering
	\begin{tabular}{|c|c|c|}
		\hline 
		Parameter & Symbol & Value  \\ 
		\hline
		\hline 
		System size & $N$ & $5$  \\
		\hline 
		Final queens interaction energy & $U_\mathrm{Q}$ & $J$  \\ 
		\hline 
		Final excluded diagonals penalty & $U_\mathrm{D}$ & $5J$  \\ 
		\hline 
		Final trapping energy & $U_\mathrm{T}$ & $2J$ \\
		\hline
		Sweep time & $\tau$ & $49 \hbar/J$ \\ 
		\hline 
		Excluded sum-diagonals & & $\{ 2,3,6,9\}$  \\ 
		\hline 
		Excluded difference-diagonals & & $\{1,2,8,9\}$ \\ 
		\hline 
		Trapping sites & & $\{ (3,5) \}$ \\
		\hline 
		Number of modes per direction & $M$ & 5 \\
		\hline
		
	\end{tabular} 
	\caption{Parameters of the exemplary instance used in the figures in the main text.}
	\label{tbl:Values}
\end{table}

We now describe how we choose the parameters used in our example. For this we calculate the minimal gap and the overlap with the final solution for several parameters to find a region with large minimal gap and large overlap. Note that this is only done to find good parameters for our small example, where we already know the solution. For large systems such a calculation would beyond classical numerical capabilities, which is why the problem poses a potential application for a quantum simulator.

The minimal gap in the spectrum (e.g.\ the one shown in Fig.\ \ref{fig:spectrum}) depends on the final queens interaction energy $U_\mathrm{Q}$, the final trapping energy $U_\mathrm{T}$, the tunneling amplitude $J$ and the final excluded diagonals penalty $U_\mathrm{D}$. To find proper values for these parameters we determine the minimal gap in a wide parameter range. In order to get the minimal gap, some of the Hamiltonian's lowest eigenenergies are calculated for discrete time steps during the sweep. Subsequently, the minimum of the difference between the groundstate and the first exited state at all time steps is taken to be the minimal gap. The values of the minimal gap have to be scrutinized carefully since its accuracy depends on the resolution of the discrete time steps. Therefore a more detailed analysis of the minimal gap might require a more careful analysis, especially for high interaction strengths.

To analyze how well the quantum system reproduces the solution of the $N$-queens problem we study the overlap  
\begin{equation}\label{eq:fidelity}
F=|\braket{\phi | \psi}|
\end{equation}
between the state $\ket{\phi}$ that corresponds to the solution of the chosen instance of the queens problem introduced in Fig.\ \ref{fig:fig1} and the state at the end of an adiabatic sweep $\ket{\psi}$ (i.e.\ the ground state of our spectrum on the right side). This is necessary because we do not switch off the kinetic Hamiltonian in our example, and thus the "perfect" solution is only obtained in the limit of large energy penalties $U_\mathrm{Q}$, $U_\mathrm{D}$ and $U_\mathrm{T}$.

\begin{figure}[tb]
	\centering
	\includegraphics{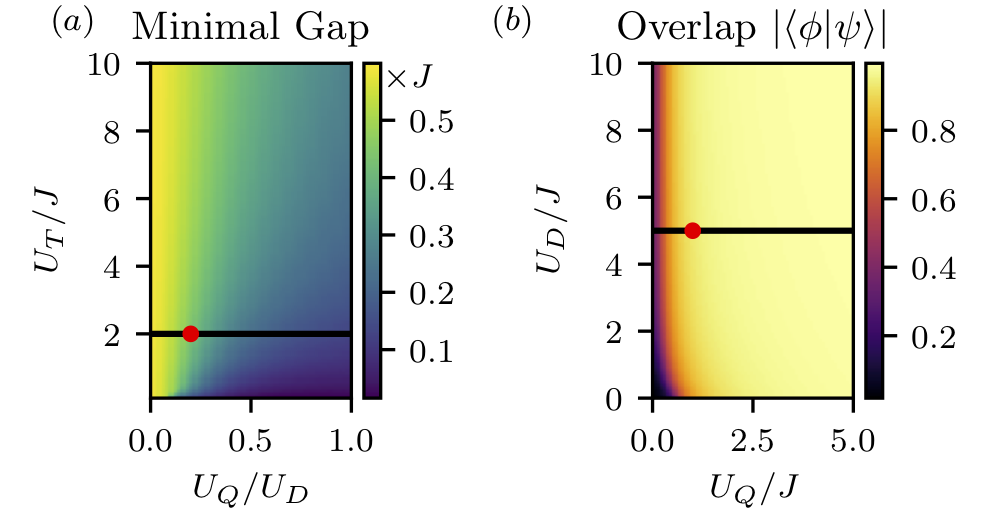}
	\caption{\textit{Discussion of the impact of parameter values.} - (a) The minimal gap of the energy spectrum for the sweep as a function of the final parameters. $U_\mathrm{D}$ is fixed to $5J$ and $U_\mathrm{Q}$ is varied. For the discussion on fidelity [subplot (b)] we choose $U_\mathrm{T} = 2J$ which is indicated by the black line. The values of $U_\mathrm{T}$ are non-zero to avoid degenerate ground states. We choose an instance with parameter values denoted by the red dot. (b) The overlap defined in Eq.\ \eqref{eq:fidelity}. The black line indicates the fixed value of $U_\mathrm{D}$ for subplot (a) and the red dot indicates the chosen set of parameters as before.}
	\label{fig:optimization}
\end{figure}

Figure \ref{fig:optimization}(a) suggests that in order to increase the minimal gap the ratio $U_\mathrm{Q}/U_\mathrm{D}$ has to be chosen as small as possible. We  vary the ratio by fixing $U_\mathrm{D}$ and varying $U_\mathrm{Q}$. Therewith, Fig.\ \ref{fig:optimization}(a) indicates that $U_\mathrm{Q}$ should be as small as possible. However, as it can be seen in Fig.\ \ref{fig:optimization}(b), a small $U_\mathrm{Q}$ also decreases the overlap with the solution and the physical system does not resemble the desired solution of the queens problem anymore. We therefore have to make a compromise between a reasonably large overlap and an optimized minimal gap.

If we set $U_\mathrm{D}$ to $5J$ and $U_\mathrm{T}$ to $2J$ we find that for $U_\mathrm{Q}=1J$ the overlap is $F \approx 0.93$ and the minimal gap is around $0.44J$. These values were used for Figs.\ \ref{fig:timeevolution} and \ref{fig:spectrum}.

\section{Derivation of the effective Hamiltonian}
\label{app:effectiveHamiltonian}
The derivation essentially follows App.\ C in Ref.\ \cite{torggler2017quantum}, generalized to two dimensions and additionally including cavity-induced tunneling (see also e.g.\ Ref.\ \cite{maschler2008ultracold}). Including the decay of the cavity fields through the mirrors with the rates $\kappa_m$, the full open-system dynamics in the tight binding limit is given by the Lindblad equation
\begin{equation}\label{eq:master_full}
\dot \rho = -\frac{i}{\hbar} [\tilde H,\rho] + \sum_m \kappa_m (2a_m \rho a_m^\dagger - a_m^\dagger a_m \rho - \rho a_m^\dagger a_m),
\end{equation}
where $\tilde H$ is specified in Eq.\ \eqref{eq:sweepModes}. The important term for the following discussion discussion is the atom-light interaction $H_\mathrm{cav}$ given in Eq.\ \eqref{eq:cavhamiltonian}.

From this master equation we obtain the Heisenberg-Langevin equations of the cavity fields
\begin{equation}
\dot a_m = (i \tilde \Delta_{c,m} - \kappa_m) a_m - i N \eta_m \hat \Theta_m  + \sqrt{2\kappa_m}\hat \xi_m
\end{equation}
with the quantum noise operator $\hat \xi_m$ obeying $[\hat \xi_m(t), \hat \xi_m^\dagger(t')] = \delta(t-t')$.

Assuming that the cavity mode fields evolve on a much faster time scale than the atomic motion ($J/(\hbar |\tilde \Delta_{c,m} + i \kappa_m|) \ll 1$), they can be approximated by their steady state on a course grained time scale \cite{torggler2017quantum,nagy2009nonlinear,habibian2013bose,maschler2008ultracold}. From the Heisenberg-Langevin equation, to zeroth order in $J/(\hbar |\tilde \Delta_{c,m} + i \kappa_m|)$, we get
\begin{equation}
a_m^\mathrm{st} \equiv \frac{\eta_m}{\tilde \Delta_{c,m} + i \kappa_m} N \hat \Theta_m + \sqrt{\frac{2\kappa_m}{\tilde \Delta_{c,m}^2+\kappa_m^2}} \hat \xi_m
\end{equation}
with $\hat \Theta_m = \frac{1}{N}\sum_{i,j=1}^N \left( v_m^{ij} \hat n_{ij} + u_m^{ij} \hat B_{ij} \right)$. That is, at steady state the effect of the field can be expressed by atomic operators only.

We now substitute the cavity field operators by their corresponding steady-state approximations in the Heisenberg equation of the atomic annihilation operators
\begin{equation}
\begin{aligned}
\dot b_{ij} = \frac{1}{i\hbar} [b_{ij}, H_\mathrm{cav}] + ... =& -i \sum_m \frac{(N\eta_m)^2}{\tilde \Delta_{c,m}^2+\kappa_m^2} \left[ \tilde \Delta_{c,m} \left( [b_{ij}, \hat \Theta_m^\dagger] \hat \Theta_m + \hat \Theta_m^\dagger [b_{ij}, \hat \Theta_m] \right) \right. \\
& \hspace*{3.2cm} \left. - i \kappa_m \left( [b_{ij}, \hat \Theta_m^\dagger] \hat \Theta_m - \hat \Theta_m^\dagger [b_{ij}, \hat \Theta_m] \right)\right] \\
& -i \sum_m \frac{N \eta_m \sqrt{2\kappa_m}}{\sqrt{\tilde \Delta_{c,m}^2+\kappa_m^2}} \left( [b_{ij}, \hat \Theta_m^\dagger] \hat \xi_m + \hat \xi_m^\dagger [b_{ij}, \hat \Theta_m] \right)
\end{aligned}
\end{equation}
where we only report terms including the cavity. At this point, ordering of atomic and field operators becomes important, since $a_m^\mathrm{st} \propto \hat \Theta_m$ as opposed to $a_m$ does not necessarily commute with atomic operators. Here we choose normal ordering, as already done in Eq.\ \eqref{eq:cavhamiltonian}. The expression contains coherent terms proportional to $\tilde \Delta_{c,m}$ and incoherent terms proportional to $\kappa_m$.

For $|\tilde \Delta_{c,m}| \gg \kappa_m$ we can neglect the incoherent part and the Heisenberg equation can be obtained from
\begin{equation}
\dot b_{ij} = \frac{1}{i\hbar} [b_{ij}, H_\mathrm{cav}^\mathrm{eff}] + ...
\end{equation}
Thus the dynamics in the coherent regime is described by the effective Hamiltonian $H_\mathrm{cav}^\mathrm{eff}$ given in Eq.\ \eqref{eq:quantumCavHamiltonian}. Otherwise the Heisenberg equation is equivalent to the master equation \eqref{eq:master}.

The results can also be obtained by naively substituting $a_m$ with $a_m^\mathrm{st}$ directly in the Hamiltonian Eq.\ \eqref{eq:cavhamiltonian} or the master equation \eqref{eq:master_full} with the given ordering.

Note that the effective Hamiltonian can also be written in the form
\begin{equation}
H_\mathrm{cav}^\mathrm{eff}  = \hbar \sum_m \tilde \Delta_{c,m} (a_m^\mathrm{st})^\dagger a_m^\mathrm{st},
\end{equation}
which allows for a simple interpretation: For $\tilde \Delta_{c,m}>0$ the lowest energy states tend to minimize the intensity of the cavity fields $\langle (a_m^\mathrm{st})^\dagger a_m^\mathrm{st} \rangle$.

\section{Harmonic approximation of potential wells}
\label{app:deeplatticelimit}
In this section we investigate the limit of a deep lattice in more detail. In Section \ref{ssec:tight-binding} we presented results in the "infinitely" deep lattice limit, where the Wannier functions become delta functions. To gain more insight to deep but finite lattice depths, we use a harmonic approximation for the potential wells. The ground state wave function is then an approximation to the lowest-band Wannier function
\begin{equation}
w_\mathrm{har}(x) = \pi^{-\frac{1}{4}} a_0^{-\frac{1}{2}}  e^{-\frac{x^2}{2a_0^2}}
\end{equation}
with the size $a_0 = (E_\mathrm{R}/V_\mathrm{L})^{1/4}/k_\mathrm{L}$ \cite{jaksch1998cold}.

With this the atom-mode overlap integrals [Eq.\ \eqref{eq:modeoverlaps}] can be calculated analytically using running wave mode functions [Eq.\ \eqref{eq:runningwaves}]. For the on-site term we obtain
\begin{equation}
v_m^{ij} = h_m(x_i,y_j) e^{-\left(\frac{k_m^x}{2k_\mathrm{L}}\right)^2\sqrt{\frac{E_\mathrm{R}}{V_\mathrm{L}}}}.
\end{equation}
It consists of the mode function at the lattice site and an exponential which reduces the overlap due to Gaussian smoothing of the mode function. As intuitively expected, the smoothing has a stronger effect for large mode wave numbers $k_m^x$. For $V_\mathrm{L}/E_\mathrm{R} \gg 1$, we obtain $v_m^{i,j} = h_m(x_i,y_j)$, as in the main text.

For the off-site overlaps we obtain
\begin{equation}
\begin{aligned}
u_m^{ij} =& h_m((x_i+x_{i+1})/2,y_j)e^{-\left(\frac{k_m^x}{2k_\mathrm{L}}\right)^2\sqrt{\frac{E_\mathrm{R}}{V_\mathrm{L}}}} e^{-\frac{\pi^2}{4} \sqrt{\frac{V_\mathrm{L}}{E_\mathrm{R}}}}.
\end{aligned}
\end{equation}
The overlap consists of three terms: First, it is the mode function evaluated in between the lattice sites. Second, there is again the Gaussian smoothing term as for the on-site overlap. Lastly, there is an exponential independent of the modes, which comes from the overlap of the two Gaussians. It goes to zero for $V_\mathrm{L}/E_\mathrm{R} \gg 1$, leading to $u_m^{ij} = 0$.

The order operator is then
\begin{equation}
\hat \Theta_m^\mathrm{har} = \frac{1}{N} e^{-\left(\frac{k_m^x}{2k_\mathrm{L}}\right)^2\sqrt{\frac{E_\mathrm{R}}{V_\mathrm{L}}}}  \sum_{i,j=1}^N \left( h_m(x_i,y_j) \hat n_{ij} + h_m((x_i+x_{i+1})/2,y_j) \hat B_{ij} e^{-\frac{\pi^2}{4} \sqrt{\frac{V_\mathrm{L}}{E_\mathrm{R}}}} \right)
\end{equation}
leading to an interaction Hamiltonian [Eq.\ \eqref{eq:quantumCavHamiltonian}] given by
\begin{equation}
\begin{aligned}
H_\mathrm{cav}^\mathrm{har} =& U_Q \sum_m f_m e^{-2 \left(\frac{k_m^x}{2k_\mathrm{L}}\right)^2\sqrt{\frac{E_\mathrm{R}}{V_\mathrm{L}}}} \sum_{ijkl} \left( h_m^*(x_i,y_j) \hat n_{ij} + h_m^*((x_i+x_{i+1})/2,y_j) \hat B_{ij} e^{-\frac{\pi^2}{4} \sqrt{\frac{V_\mathrm{L}}{E_\mathrm{R}}}} \right) \\
& \hspace*{4.15cm} \times \left( h_m(x_k,y_l) \hat n_{kl} + h_m((x_k+x_{k+1})/2,y_l) \hat B_{kl} e^{-\frac{\pi^2}{4} \sqrt{\frac{V_\mathrm{L}}{E_\mathrm{R}}}} \right),
\end{aligned}
\end{equation}
which in the "infinitely" deep lattice limit simplifies to Eq.\ \eqref{eq:quantumCavHamiltonianDL}. All cavity-induced tunneling terms are suppressed by the exponential and tend to be smaller than density-density terms. Also, since $U_\mathrm{Q}$ is maximally on the order of $J$ (at the end of the sweep), cavity-induced tunneling terms are smaller than $H_\mathrm{kin}$. However, also the density-density terms can be small for example when $h_m(x_i,y_j)=0$, which is why we still include cavity-induced tunneling in the simulations.

In the main text we chose uniform $f_m = 1/M$. To compensate for Gaussian smoothing one might want to include the exponential as correction
\begin{equation}
\tilde f_m = f_m e^{2\left(\frac{k_m^x}{2k_\mathrm{L}}\right)^2\sqrt{\frac{E_\mathrm{R}}{V_\mathrm{L}}}},
\end{equation}
which leads to even better results ($\tilde A$ is closer to $A$ for finite lattice depths). Note that this correction does only depend on $V_\mathrm{L}$ and not on the problem size or number of modes in our implementation, since the range of $k_m^x$ is fixed.

\section{Shape of the interaction}
\label{app:interaction}

Here we reformulate the interaction in the infinitely deep lattice limit from Eq.\ \ref{eq:intfunc} with Fourier transforms by defining a real envelope function $f(k)$ such that $f(k_m) = f_m$. To get back the discrete wave numbers, this function is sampled with a Dirac comb at the lines $m\Delta k + k_s$ with $m \in \mathbb{Z}$, where $k_s$ is a constant shift and $\Delta k$ is the spacing between the pumped modes. In the main text we only consider the case $k_s = 0$ for simplicity. We define the Fourier transform as $\mathcal{F}\{f\}(r) = \int_{-\infty}^\infty \mathrm dk f(k) e^{ikr}/\sqrt{2\pi}$ and denote the convolution as $(f*g)(t) = \int_{-\infty}^\infty f(\tau) g(t-\tau) \mathrm d \tau$.

For simplicity, we take parallel wave vectors $\boldsymbol k_m$. Along this direction $\boldsymbol r \parallel \boldsymbol k_m$ we write
\begin{equation}
\begin{aligned}
\tilde{\mathcal{A}}(r) &= \sum_m f_m \cos(k_m r) = \int_{-\infty}^\infty \mathrm dk \, f(k) \sum_{m=-\infty}^\infty \delta(k-m\Delta k-k_s) \cos(kr)\\
&= \mathrm{Re} \left[ \sqrt{2\pi} \mathcal{F} \left\{ f(k) \sum_{m=-\infty}^\infty \delta(k - m \Delta k-k_s) \right\}(r) \right] = \mathrm{Re}\left[\sqrt{2\pi} \mathcal{F}\{f\}(r) * \sum_{l=-\infty}^\infty \delta\left(r - l \frac{2\pi}{\Delta k}\right) e^{i k_s r} \right]\\
&= \sum_{l=-\infty}^\infty \mathrm{Re}\left[\sqrt{2\pi} \mathcal{F}\{f\}\left(r - l \frac{2\pi}{\Delta k}\right) e^{i2\pi l \frac{k_s}{\Delta k}} \right],
\end{aligned}
\end{equation}
where $r = |\boldsymbol r|$. We used the convolution and shift theorem from Fourier analysis in the second to last line and evaluated the convolution integrals by pulling out the sum in the last line.

For symmetric envelopes centered around $k_c$ we can further simplify using the shift theorem
\begin{equation}
\tilde{\mathcal{A}}(r) = \sum_{l=-\infty}^\infty \sqrt{2\pi} \mathcal{F}\{\tilde f\}\left(r - l \frac{2\pi}{\Delta k}\right) \cos\left(k_c \left(r - l \frac{2\pi}{\Delta k}\right) + 2\pi l \frac{k_s}{\Delta k}\right),
\end{equation}
where $\tilde f(k) = f(k+k_c)$ is the shifted envelope centered around $k=0$, whose Fourier transform is real.

Let us apply this to our example described in Section \ref{subsec:nqueensinteraction} and find out why the interaction has the desired property given in equation Eq.\ \eqref{eq:combproperty}. There we had uniform $f_m = 1/M$ and wave numbers
\begin{equation}
k_m^0= k_\mathrm{L} \left(1 + \frac{2m+1}{2M}\right)
\end{equation}
with $m = 0,...,M-1$. These have a mode spacing of $\Delta k = k_\mathrm{L}/M$ and are centered around $k_c = 3k_\mathrm{L}/2$. One can see that only the odd modes of the cavity (wave numbers $k_n = n \Delta k_\mathrm{FSR}$ with $n$ odd and free spectral range $\Delta k_\mathrm{FSR}$) are used. Therefore, $\Delta k = 2\Delta k_\mathrm{FSR}$ and $k_s = \Delta k / 2$, because the comb has to be shifted to fit the odd modes. Due to the uniform $f_m$ the envelope is a rectangular function with width $k_\mathrm{L}$ and height $1/M$ centered at $k_c$
\begin{equation}
f(k) = \mathrm{rect}((k-k_c)/k_\mathrm{L})/M =
\begin{cases}
1/M & \mathrm{for} \; k \in [k_c-k_\mathrm{L}/2, k_c+k_\mathrm{L}/2]\\
0 & \mathrm{otherwise}.
\end{cases}
\end{equation}

The Fourier transform of a rectangular function centered around zero with unit width and height is a sinc function $\mathrm{sinc}(x) = \sin(x)/x$. Using the addition theorem for the cosine and noting that $\cos(\pi l) = (-1)^l$ and $\sin(\pi l) = 0$ we obtain an analytical expression for the interaction
\begin{equation}
\tilde{\mathcal{A}}(r) = \sum_{l=-\infty}^\infty (-1)^l \mathrm{sinc}\left(\frac{k_\mathrm{L}}{2} \left(r - l \frac{2\pi}{\Delta k}\right)\right) \cos\left(k_c \left(r - l \frac{2\pi}{\Delta k}\right)\right).
\end{equation}
For $l=0$ and at lattice site spacings $r_j = j \pi / k_\mathrm{L}$ it takes the values
\begin{equation}
\mathrm{sinc}(\pi j/2) \cos(3\pi j/2) =
\begin{cases}
1 \quad \mathrm{for} \; j = 0\\
0 \quad \mathrm{otherwise},
\end{cases}
\end{equation}
as desired. This comes from well known properties of sinc and cosine
\begin{align*}
\mathrm{sinc}(\pi j/2) &=
\begin{cases}
1 &\mathrm{for} \; j = 0\\
0 &\mathrm{for} \; j \; \mathrm{even}\\
(-1)^\frac{j-1}{2} \frac{2}{\pi j} &\mathrm{for} \; j \; \mathrm{odd}\\
\end{cases}\\
\cos(3\pi j/2) &=
\begin{cases}
(-1)^\frac{j}{2} \quad &\mathrm{for} \; j \; \mathrm{even}\\
0 \quad &\mathrm{for} \; j \; \mathrm{odd}.
\end{cases}
\end{align*}
The other summands have the same form, but are shifted by $R = 2\pi/\Delta k = 2 \pi M / k_\mathrm{L} = r_{2M}$ ($2M$ lattice sites) and have alternating signs. Since $R$ is an integer multiple of the lattice spacing this adds up to the desired interaction given in Eq.\ \eqref{eq:combproperty} in the main text.

Thus for rectangle envelopes the bandwidth $\Delta k_\mathrm{BW}$ determines the zeros of the interaction. Taking $\Delta k_\mathrm{BW} = 2k_\mathrm{L}$ would lead to zeros at all lattice sites. For the smaller bandwidth $\Delta k_\mathrm{BW} = k_\mathrm{L}$ used here, only even sites become zero. This can be compensated by choosing a central wave number $k_c = n k_\mathrm{L}/2$ with $n$ odd, which is responsible for the zeros at odd sites. The mode spacing $\Delta k$ determines the peak distance. Finally, using odd cavity modes (specifying $k_s$) leads to alternating peaks, which does not have an effect in our implementation, since $-2M < j < 2M$.

\end{document}